\documentclass[11pt]{article}

\usepackage{geometry}
\usepackage[numbers]{natbib}
\usepackage{amsfonts,amsmath,amssymb,amsthm}
\usepackage{graphicx}
\usepackage[colorlinks,citecolor=green,linkcolor=blue,bookmarks=false,hypertexnames=true]{hyperref}
\usepackage{mathrsfs}
\usepackage{algorithm}
\usepackage{algpseudocode}
\usepackage{tikz}
\usepackage{xcolor}

\newtheorem{theorem}{Theorem}

\newtheorem{definition}{Definition}

\newtheorem{assumption}{Assumption}

\algnewcommand{\IfThenElse}[3]{% \IfThenElse{<if>}{<then>}{<else>}
  \State \algorithmicif\ #1\ \algorithmicthen\ #2\ \algorithmicelse\ #3}

\usepackage{titling}
\usepackage{array}
\preauthor{\begin{center}
    \large \lineskip .75em%
        \begin{tabular}[t]{>{\centering\arraybackslash}p{.45\textwidth}}}
        \postauthor{\end{tabular}\par\end{center}}
\makeatletter
\renewcommand\and{% \begin{tabular}
  \end{tabular}%
  \hfill
  \begin{tabular}[t]{>{\centering\arraybackslash}p{.45\textwidth}}}% \end{tabular}
\makeatother

\def\sarg{\emph{Sargassum}}

\def\ebomb{eBOMB}
\def\jsd{\textnormal{JSD}}

\newcommand{\neigh}[1]{\text{neighbor}(#1)}
\newcommand{\namedref}[2]{\hyperref[#1]{#2}}
\DeclareMathOperator*{\argmax}{arg\,max}

\allowdisplaybreaks
\usepackage{framed}
\definecolor{shadecolor}{RGB}{204,229,255}

\begin{document}

\title{Charting the course of \emph{Sargassum}: Incorporating nonlinear elastic interactions and life cycles in the Maxey--Riley model}

\author{
Gage\ Bonner\thanks{Corresponding author.} \\ Department of Atmospheric Sciences\\ Rosenstiel School of Marine, Atmospheric \& Earth Science\\ University of Miami\\ Miami, Florida, USA\\ gbonner@miami.edu
\and
Francisco J.\ \ Beron-Vera\\ Department of Atmospheric Sciences\\ Rosenstiel School of Marine, Atmospheric \& Earth Science\\ University of Miami\\ Miami, Florida, USA\\ fberon@miami.edu 
\and
Maria J.\ \ Olascoaga\\ Department of Ocean Sciences\\ Rosenstiel School of Marine, Atmospheric \& Earth Science\\ University of Miami\\ Miami, Florida, USA\\ jolascoaga@miami.edu} 

\date{Started: April 5, 2024. This version: \today. To appear in \emph{P\hspace{-.025cm}N\hspace{-.025cm}A\hspace{-.025cm}S Nexus}.\vspace{-0.25in}}

\maketitle

\begin{abstract}
The surge of pelagic \emph{Sargassum} in the Intra-America Seas, particularly the Caribbean Sea, since the early 2010s has raised significant ecological concerns. This study emphasizes the need for a mechanistic understanding of \emph{Sargassum} dynamics to elucidate the ecological impacts and uncertainties associated with blooms. By introducing a novel transport model, physical components such as ocean currents and winds are integrated with biological aspects affecting the \emph{Sargassum} life cycle, including reproduction, grounded in an enhanced Maxey--Riley theory for floating particles. Nonlinear elastic forces among the particles are included to simulate interactions within and among \emph{Sargassum} rafts. This promotes aggregation, consistent with observations, within oceanic eddies, which facilitate their transport. This cannot be achieved by the so-called leeway approach to transport, which forms the basis of current \emph{Sargassum} modeling. Using satellite-derived data, the model is validated, outperforming the leeway model. Publicly accessible codes are provided to support further research and ecosystem management efforts. This comprehensive approach is expected to improve predictive capabilities and management strategies regarding \emph{Sargassum} dynamics in affected regions, thus contributing to a deeper understanding of marine ecosystem dynamics and resilience.

\begin{snugshade*}
\begin{center}
\textbf{Significance Statement}
\end{center}

The massive increase in \emph{Sargassum} seaweed in the Caribbean Sea since the early 2010s poses major ecological challenges. This research introduces a new model that integrates ocean currents, winds, and biological factors to better understand and predict \emph{Sargassum} movement. Unlike existing models, this new approach simulates interactions within \emph{Sargassum} rafts and their aggregation in ocean eddies, providing a more accurate prediction of their movement. Validated with satellite data, the model surpasses current methods, offering improved tools for researchers and policymakers. The model is distributed in a fully open source manner and features an interface that enables use by non-experts.
\end{snugshade*}
\end{abstract}

\begin{quote}
    \emph{``Just before it was dark, as they passed a great island of Sargasso weed that heaved and swung in the light sea as though the ocean were making love with something under a yellow blanket, \dots{}.''}
    
    \flushright{Ernest Hemingway, \emph{The Old Man and the Sea}.}
\end{quote}

\section{Introduction}
\label{sec:introduction}

The brown macroalgae \emph{Sargassum} was reportedly first documented by Christopher Columbus and his crew in 1492 \cite{Godinez-etal-21}. The floating rafts, characterized by small gas-filled bladders, apparently reminded them of ``salgazo,'' a grape variety found in Portugal. Originally, abundant throughout the North Atlantic subtropical gyre, known as the Sargasso Sea. Subsequently, beginning in the early 2010s, massive pelagic \emph{Sargassum} rafts began to inundate the Intra-America Seas, notably the Caribbean Sea, during the spring and summer months \citep{Langin-18}.

These pelagic \emph{Sargassum} rafts provide essential habitats for a diverse range of marine fauna, significantly enhancing the region's biodiversity \citep{Bertola-etal-20}. Additionally, they act as carbon sinks, contributing to the global carbon cycle \citep{Paraguay-etal-20}. However, these benefits come with notable drawbacks. The rafts accumulate toxic substances and heavy metals. Upon entering coastal zones, they can cause the mortality of fish and sea turtles and suffocate vital seagrass and coral communities \citep{vanTussenbroek-etal-17}. The breakdown of extensive onshore \emph{Sargassum} rafts emits hydrogen sulfide, which may present health hazards to people. Furthermore, the mass stranding of algae on beaches significantly reduces tourism, disrupting the local economy \citep{Smetacek-Zingone-13, Resiere-etal-18}.

Comprehensive analyses of satellite imagery have revealed a persistent band of high-density \emph{Sargassum} across the tropical North Atlantic \citep{Gower-etal-13, Wang-etal-19}. The factors that precipitate blooms of pelagic \emph{Sargassum} remain a subject of active scientific debate and inquiry, as do the factors that maintain its presence across this region \citep{Wang-etal-19, Johns-etal-20, Lapointe-etal-21}. Currently, the most significant source of uncertainty is the fundamental understanding of the movement of pelagic \emph{Sargassum} \citep{Putman-etal-18,  Jouanno-etal-21a, Jouanno-etal-21b, Jouanno-etal-23}. Without a mechanistic understanding of \emph{Sargassum} dynamics, it is challenging to assess how much of the observed distributions result from transport processes versus local growth, and how these distributions will evolve over time \citep{Putman-etal-18, Putman-18a, Putman-18b}.

Several studies have attempted to account for the spatiotemporal variability in its distribution by tracking synthetic fluid particles within ocean circulation model velocity fields. Results vary significantly depending on whether simulations assume winds contribute to \emph{Sargassum} movement (e.g., through a windage or Stokes drift factor) and on assumptions about growth and mortality (typically accounted for in models by how long particles are tracked). Some studies indicate evidence of occasional transport from the Sargasso Sea into the tropics, while others suggest this to be unlikely \citep{Franks-etal-16, Wang-etal-19, Johns-etal-20}.

To address the uncertainties associated with the largely piecemeal and ad-hoc approaches in \emph{Sargassum} modeling so far, this paper develops a mechanistic model that integrates two critical aspects: one physical and one biological. The physical aspect involves transport resulting from the combined action of ocean currents and winds, mediated by \emph{inertial} effects. The biological aspect encompasses \emph{Sargassum} reproduction through \emph{vegetative growth and fragmentation} and \textcolor{black}{death through factors such as aging and sinking, mediated by general mortality.}

The modeling of inertial effects builds on the \emph{Maxey--Riley theory} for the motion of finite-size, buoyant (i.e., \emph{inertial}) particles immersed in fluid flow \citep{Maxey-Riley-83, Auton-etal-88, Michaelides-97}. The Maxey--Riley equation represents an application of Newton's second law, \textcolor{black}{incorporating various forces, including the flow force exerted on the particle by the undisturbed fluid, the added mass force resulting from part of the fluid moving with the particle, and the drag force caused by the fluid viscosity, among others. These forces are derived from \emph{first principles}, as they originate by integrating the Cauchy stress tensor over the spherical surface of the particle immersed in a moving viscous fluid, assuming the Stokes number is small \citep{Maxey-Riley-83}. For a turbulent flow such as the ocean flow, this imposes a restriction on the particle size; depending on the chosen scales, its radius should be (much) smaller than about 1 meter.} 

The motion of inertial particles thus deviates fundamentally from that of infinitesimally small, neutrally buoyant (i.e., fluid) particles, as inertial particles are incapable of adjusting their velocities to conform to the flow velocity \cite{Cartwright-etal-10}. Although the trajectories of neutrally buoyant particles eventually synchronize with those of fluid particles over time, they do so relative to fluid particles starting from different initial positions \cite{Babiano-etal-00, Sapsis-Haller-08}. The manifestation of inertial effects, which have been documented even for minute neutrally buoyant particles \cite{Sapsis-etal-11}, \textcolor{black}{can be expected} to be significantly magnified in the dynamic context of sprawling rafts of floating \emph{Sargassum}.

The fluid mechanics Maxey--Riley theory has been extended to particles \emph{floating at the ocean surface} \citep{Beron-etal-19-PoF}, enabling them to \emph{interact through elastic forces} \citep{Beron-Miron-20}. The oceanographic Maxey-Riley framework for individual particles has undergone successful testing in both the field \citep{Olascoaga-etal-20, Miron-etal-20-GRL} and the laboratory \citep{Miron-etal-20-PoF}, and is exhaustively reviewed in \citep{Beron-21-ND}. 

Elastically interacting particles serve as a parsimonious model for \emph{Sargassum} plant architecture, representing flexible stems as springs and air-filled bladders as inertial particles. These interactions lead to behavior that differs from noninteracting floating inertial particles, promoting concentration within mesoscale eddies independent of their polarity. This behavior is supported by rigorous results \citep{Beron-Miron-20} and empirical evidence \cite{Andrade-etal-22}.

The Maxey--Riley model for \emph{Sargassum} plant motion is significantly extended here to simulate interactions among rafts, conceptualized as networks subjected to \emph{nonlinear} elastic forces. The modeling of \emph{Sargassum} reproduction is approached mechanistically by regulating the size of the networks based on a specified law.

In a notable stride forward, this paper not only validates the model against satellite-derived Sargassum densities, outperforming the so-called leeway approach to transport that supports contemporary \emph{Sargassum} modeling \cite{Podlejski-etal-23, Jouanno-etal-23, Lara-etal-24}. Additionally, this paper develops and publicly releases codes for executing model simulations and computing densities. Consequently, the community will have access to a robust and flexible toolkit for in-depth exploration of Sargassum dynamics, enhancing understanding and predictive capabilities in \emph{Sargassum} research, management, and response efforts.

\section{Results}
\label{sec:results}

\subsection{Model overview} 
\label{sec:model-overview}

We delineate the terminology pertinent to directly related models. The \emph{BOM model} was introduced in \cite{Beron-etal-19-PoF} as a comprehensive Maxey--Riley framework to characterize the dynamics of floating particles of small but finite dimensions on the ocean surface. The \emph{eBOM model} was presented in \cite{Beron-Miron-20} as a model for \sarg{} plants, conceptualized as networks of elastically interacting BOM particles. In this paper, we advance the \emph{eBOMB model} as an extension of the aforementioned model, as elaborated in the subsequent sections.

\begin{figure}[t!]
    \centering
    \includegraphics[width = 0.75\textwidth]{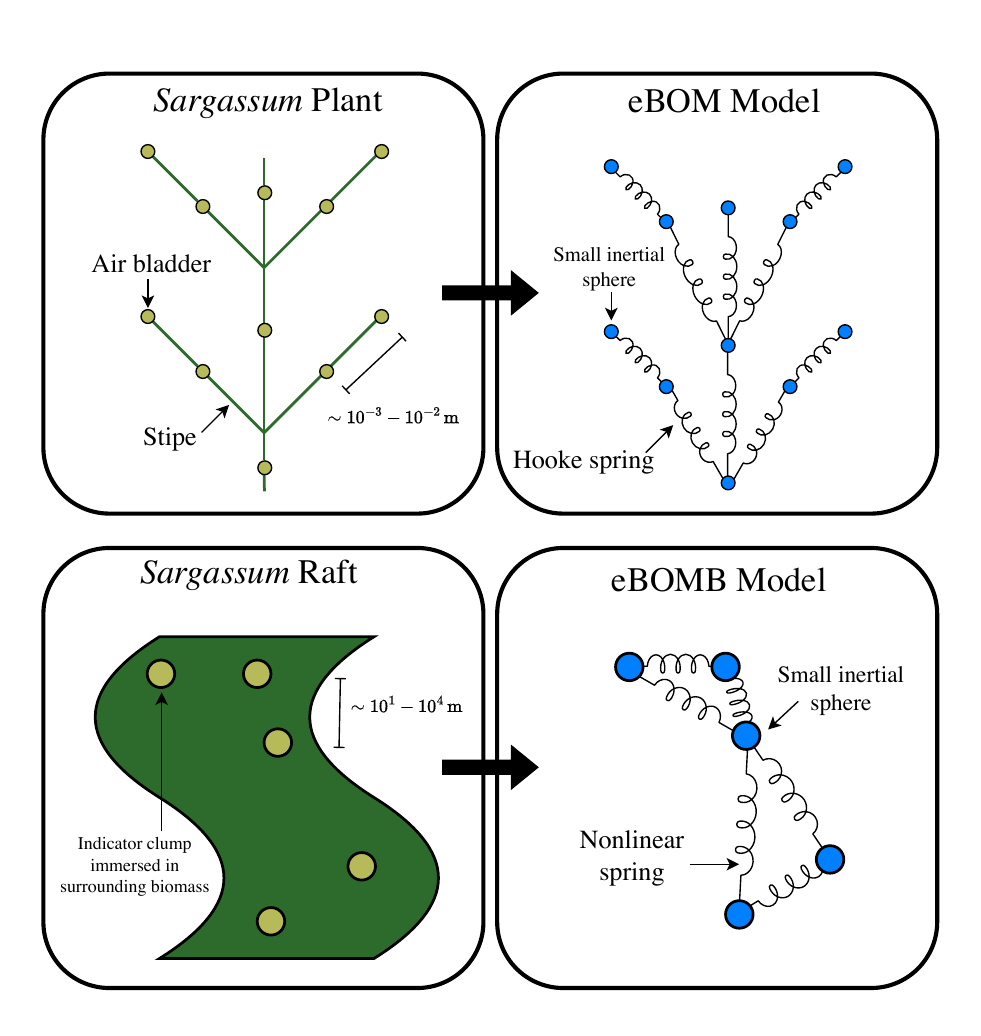}
    \caption{(top row) \textcolor{black}{As viewed from above,} an individual \sarg{} plant is idealized as a network of air bladders connected by semi-rigid stipes. The eBOM model treats this as a network of small inertial spheres, each of which obeys a Maxey-Riley equation coupled by linear elastic (Hookian) forces. (bottom row) \textcolor{black}{As viewed from above,} a \sarg{} raft is idealized as a continuous biomass dotted with \emph{indicator clumps} that approximate that raft's motion. The \ebomb{} model treats the clumps as Maxey-Riley particles similarly to eBOM, but the connecting springs have nonlinear stiffness parameters that allow for fragmentation and recombination of subrafts.}
    \label{fig:model-cartoon}
\end{figure}

The fundamental object of the \ebomb{} model is a \emph{clump} of \sarg{} in the open ocean, subject to currents and wind. A clump is qualitatively a handful of \sarg{} that we consider to be a hard sphere with a small but finite radius. We call a network of interacting clumps a \emph{raft} and assume that the motion of a continuous macroscopic quantity of \sarg{} can be well approximated by the motion of this discrete set of clumps as ``landmarks.'' This is represented schematically along with a comparison to the precursor eBOM model in Fig.\@~\ref{fig:model-cartoon}.
The totality of \sarg{} under consideration at a given time is itself a raft although it may consist of a number of disconnected networks of clumps which may be referred to as \emph{subrafts}. \textcolor{black}{The overall size hierarchy of \sarg{} quantities is therefore, from smallest to largest: air bladders, stipes, plants, clumps, subrafts, and rafts.} The model consists of three main components, the clump physics controlled by coupled Maxey--Riley equations, spring forces with nonlinear stiffness parameters connecting subrafts, and the biological model controlling the growth and death of clumps. 

\subsubsection{Physical module of the eBOMB model}

The Maxey--Riley physics follows \cite{Beron-etal-19-PoF, Beron-Miron-20}. There are four parameters required to define the main equations: $\alpha$ is the unitless parameter measuring \emph{windage}, $\tau$ is the \emph{inertial response time}, and $R$ is a dimensionless parameter that quantifies the exposure of the assumed spherical clump to the air. These parameters depend further on the radius $a$ of a clump and the density ratio of the clump to the water $\delta \ge 1$, which we refer to as \emph{buoyancy} (additional details are given in Materials and Methods). We will assume here that each of these parameters can be specified independently, although we will justify their particular values with reference to $\delta$ and $a$ later. Let $\mathbf x = (x, y)$ denote the position on the ocean surface with rescaled longitude--latitude (i.e., geographic) coordinates $x = a_\odot\cos\vartheta_0(\lambda - \lambda_0)$ and $y = a_\odot(\vartheta - \vartheta_0)$, where $a_\odot$ is the mean Earth's radius and $(\lambda_0,\vartheta_0)$ denotes a reference location. We define $\gamma_\odot(y) = \sec\vartheta_0\cos\vartheta$ and $\tau_\odot(y) = a_\odot^{-1}\tan\vartheta$, which are geometric factors that arise due to the sphericity of the Earth. Let $\mathbf v(\mathbf x, t)$ and $\mathbf v_\text W(\mathbf x, t)$ denote the near-surface ocean and wind velocities, respectively. Define
\begin{equation} \label{eq:u-def}
    \mathbf u = (1 - \alpha) \mathbf v + \alpha \mathbf v_\text W.
\end{equation}
We decompose $\mathbf v = \mathbf v_\text E + \sigma \mathbf v_\text S$, where $\mathbf v_\text E$ represents the part of the ocean velocity, sometimes referred as ``Eulerian,'' on time scales much longer than the period of surface gravity waves and $\mathbf v_\text S$ is the wave-induced Stokes drift.  This decomposition follows \cite{Craik-82}, except for the unitless quantity $\sigma$, which we introduce to parameterize uncertainty around $\mathbf v_\text S$. Let $\omega = \gamma_\odot^{-1}\partial_x v_y - \partial_y v_x + \tau_\odot v_x$ be the vorticity of the ocean velocity and let $\nabla_{\mathbf w}\mathbf w = \gamma_\odot^{-1}w_x\partial_x \mathbf w + w_y\partial_y \mathbf w +  \tau_\odot w_x\mathbf w^\perp$ be the covariant derivative of $\mathbf w = \mathbf v$ or $\mathbf u$ in the direction of itself, where $\perp$ stands for anticlockwise $\frac{\pi}{2}$-rotation.  The trajectory $\mathbf x_i(t)$ of the $i$th clump of a raft obeys (see Supplementary Materials for details):
\begin{subequations}\label{eq:MR-high-level-def}
\begin{equation} 
    \sqrt{\mathsf m}\,\dot{\mathbf x}_i = \mathbf u\vert_i + \tau \mathbf u_\tau\vert_i  +  \tau \mathbf F_i,
\end{equation}
where
\begin{equation}
    \mathbf u_\tau = R\big(\partial_t\mathbf v + \nabla_{\mathbf v}\mathbf v\big) + R \left( f + \tfrac{1}{3}\omega \right)\mathbf v^\perp -  \partial_t\mathbf u - \nabla_{\mathbf u}\mathbf u- \left( f + \tau_\odot u_x + \tfrac{1}{3}R \omega \right) \mathbf u^\perp.
\end{equation}
\end{subequations}
In this context, the matrix $\mathsf m$ is the representation of metric in the (rescaled geographic) coordinates $(x,y)$, with elements $\mathsf m_{11} = \gamma_\odot^2$, $\mathsf m_{11} = 1$ and $\mathsf m_{12} = 0 = \mathsf m_{21}$. The notation $\vert_i$ signifies the association with clump $i$. The function $f(y) = 2\Omega\sin\vartheta$, where $\Omega$ denotes the Earth's rotation rate, is the Coriolis parameter. Additionally, $\mathbf F_i$ denotes an external force. Consequently, for a raft consisting of $N$ clumps, we obtain a system of $2N$ first-order ordinary differential equations, \emph{which are coupled via $\mathbf F_i$}. 

We now describe the interaction spring forces. Let $\mathbf F_i$ be the force felt by the clump labeled $i$ and let $\mathbf x_{ij} = \mathbf x_i - \mathbf x_j$, that is,
\begin{equation} \label{eq:F-spring-def}
    \mathbf F_i = - \sum_{j \in \neigh{i, t}} k(|\mathbf x_{ij}|) (|\mathbf x_{ij}| - L) \frac{\mathbf x_{ij}}{|\mathbf x_{ij}|},
\end{equation}
where $|\mathbf x_{ij}|^2 = \gamma_\odot^2(x_i-x_j)^2 + (y_i-y_j)^2$. Additionally, $k(|\mathbf x_{ij}|)$ represents the \emph{stiffness function}, $L$ is the \emph{natural length} of the spring, and $\neigh{i, t}$ is equal to the set of indices of clumps that are connected to $i$ at time $t$. Note that all springs have the same $k$ and $L$. The motivation behind this force is the effect of inter-clump entanglement which we assume provides a restorative force up to a certain distance, at which the clumps dissociate completely. Hence, we write
\begin{equation} \label{eq:k-bomb-def}
    k(|\mathbf x_{ij}|) = \frac{A}{\mathrm{e}^{(|\mathbf x_{ij}| - 2L)/\Delta} + 1},
\end{equation}
where $\Delta$ is taken small enough such that $k(|\mathbf x_{ij}|) \approx A$ for $0\le |\mathbf x_{ij}| \le 2L$ and $k(|\mathbf x_{ij}|) \approx 0$ for $|\mathbf x_{ij}| > 2 L$. It is worth mentioning that a piecewise constant $k(|\mathbf x_{ij}|)$ produces almost the same numerical results, yet Eq.\@~\eqref{eq:k-bomb-def} is a differentiable function that meets the technical conditions of Theorem \ref{thm:vortices}. We take $\neigh{i, t}$ to be a set of nearest-neighbors to clump $i$ at time $t$. This allows us to ensure that the simulations are fast by ignoring distant clumps, while also providing control over the size of subrafts, i.e., taking the number of nearest-neighbors to be large results in subrafts with greater membership. The natural length of the spring $L$ is computed based on the initial configuration of clumps so that the model scales appropriately. In particular, if $d_i(K)$ is the mean distance of clump $i$'s $K$ nearest-neighbors at the initial integration time, then $L$ is chosen to be the median among all $d_{i}(K)$.

\subsubsection{Biological module of the eBOMB model}

To simulate \sarg{} life cycles, we employ a hybrid version of the models proposed by \cite{Brooks-etal-19} and \cite{Jouanno-etal-21b}, specifically adapted to the scenario of a discrete set of clumps as delineated below. Each clump is allocated an \emph{amount} parameter, denoted in arbitrary units as $S(t)$, and is initialized with the condition $S(0) = S_0$. At every integration time step $h$ of the coupled system defined by Eq.\@~\eqref{eq:MR-high-level-def}, $S(t)$ is updated via
\begin{equation} \label{eq:S-clump-amount-def}
    S(t + h) = S(t) + (g(t) - m) h, 
\end{equation}
where $g(t)$ and $m$ represent the factors controlling growth and mortality, respectively. We take $m$ to be a constant and $g(t)$ equal to a constant multiplied by factors depending on the temperature $T$ and nitrogen content of the water $N$ via
\begin{subequations}\label{eq:life}
\begin{equation}
    g(t) = \mu_\text{max} \cdot \frac{\mathcal{T}(t)}{k_\text{N}/N(t) + 1},
\end{equation}  
where
\begin{equation}
    \mathcal{T}(t) 
    = 
    \begin{cases}
        \exp\Big(-\frac{1}{2}\left(\frac{T - T_0}{T - T_{\text{min}}}\right)^2\Big) & \text{if } T_{\text{min}} \leq T(t) \leq T_0, \\ 
        \exp\Big(-\frac{1}{2}\left(\frac{T - T_0}{T - T_{\text{max}}}\right)^2\Big) & \text{if } T_0 < T(t) \leq T_{\text{max}}, \\ 
        0 & \text{otherwise}.
    \end{cases}
\end{equation}
\end{subequations}
Here, $\mu_\text{max}$ is the maximum \sarg{} growth rate, $T_\text{min}, T_\text{max}$ are the minimum and maximum temperatures for \sarg{} growth, $T_0 = (T_\text{min} + T_\text{max})/2$ is the (approximate) optimal growth temperature and $k_\text{N}$ is the \sarg{} nitrogen uptake half saturation. Once Eq.\@~\eqref{eq:S-clump-amount-def} has been applied at the end of the time step, we compare $S(t + h)$ to thresholds $S_{\text{min}}$ and $S_{\text{max}}$. If $S(t + h) > S_{\text{max}}$, a new clump is born with $S_{\text{new}}(t + h) = S_0$. The new clump is placed a distance $L$ away from its ``parent'' at an angle chosen uniformly at random, and the parent is reset to $S(t + h) = S_0$. If $S(t + h) < S_{\text{min}}$, the clump is removed completely from the integration by dynamically reducing the size of the integrator cache. Clumps may also die when reaching land. 

\subsection{Model validation}

Our principal comparative framework employs the \emph{leeway model}, which governs the transport dynamics of \emph{Sargassum} within state-of-the-art simulations \cite{Jouanno-etal-23}.  Frequently referenced in the search-and-rescue literature \cite{Breivik-etal-13}, the leeway model is defined as
\begin{equation}\label{eq:leeway-def}
    \dot{\mathbf x}_i = \mathbf v_\text E|_i + \alpha_{10} \mathbf v_{10}|_i,
\end{equation}
where the unitless parameter $\alpha_{10}$ is chosen in an ad hoc manner, typically in the range 0.01--0.03 \cite{Jouanno-etal-23}, to measure windage, and $\mathbf v_{10}$ is the wind velocity at 10-m above the sea level. This is fundamentally different from the windage parameter $\alpha$ in the eBOMB model. Although the carrying flow velocity $\mathbf u$ in the eBOMB model, as described in Eq.\@~\ref{eq:u-def}, appears similar to the leeway model in Eq.\@~\ref{eq:leeway-def}, $\mathbf u$ is not arbitrarily chosen. Instead, it results from vertically averaging the Stokes drag exerted by the ocean velocity on the submerged part of an inertial particle and by the wind velocity on the exposed part \cite{Beron-etal-19-PoF}. For example, we will refer to Eq.~\eqref{eq:leeway-def} with $\alpha_{10} = 0.01$ as ``Leeway 1\%''. The accuracy of both the \ebomb{} and leeway models will be evaluated against \sarg{} coverage distributions obtained from satellite data (see Material and Methods for more details on how these distributions are derived).

\begin{figure}[t!]
    \centering
    \includegraphics[width = .75\textwidth]{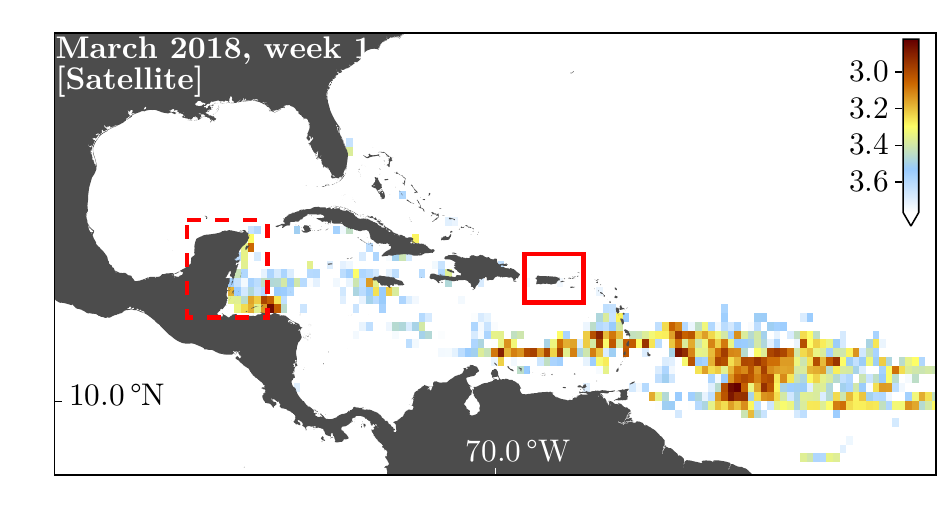}
    \caption{AFAI-derived \sarg{} coverage distribution from which the simulations in the subsequent figures were initialized. The value the colorbar is $-\log_{10}b$ where $b$ is the monthly fraction of \sarg{} in a bin. Highlighted are regions encompassing Puerto Rico Island (solid red) and the Yucatan Peninsula (dashed red), chosen for model validation.}
    \label{fig:initial-sarg}
\end{figure}

In this research, we focus our simulations on the year 2018, a time characterized by significant \sarg{} presence in the Caribbean Sea. These simulations employ Eulerian ocean velocity ($\mathbf v_\text E$) derived from an amalgamation of various observational datasets, alongside Stokes drift ($\mathbf v_\text S$) and wind velocity ($\mathbf v_\text W$) generated by a reanalysis system. The parameters for our primary simulation were determined through a synthesis of heuristics, established values, and optimization. We initialize both the \ebomb{} and leeway models based on the \sarg{} coverage distribution accumulated during the first week of March, as illustrated in Fig.\@~\ref{fig:initial-sarg}, and subsequently simulate each model over a 14-week period. Details regarding the construction of the velocity fields, parameter optimization, and the computation of \sarg{} coverage distribution are provided in Materials and Methods.

\begin{figure}[t!]
    \centering
    \includegraphics[width = \textwidth]{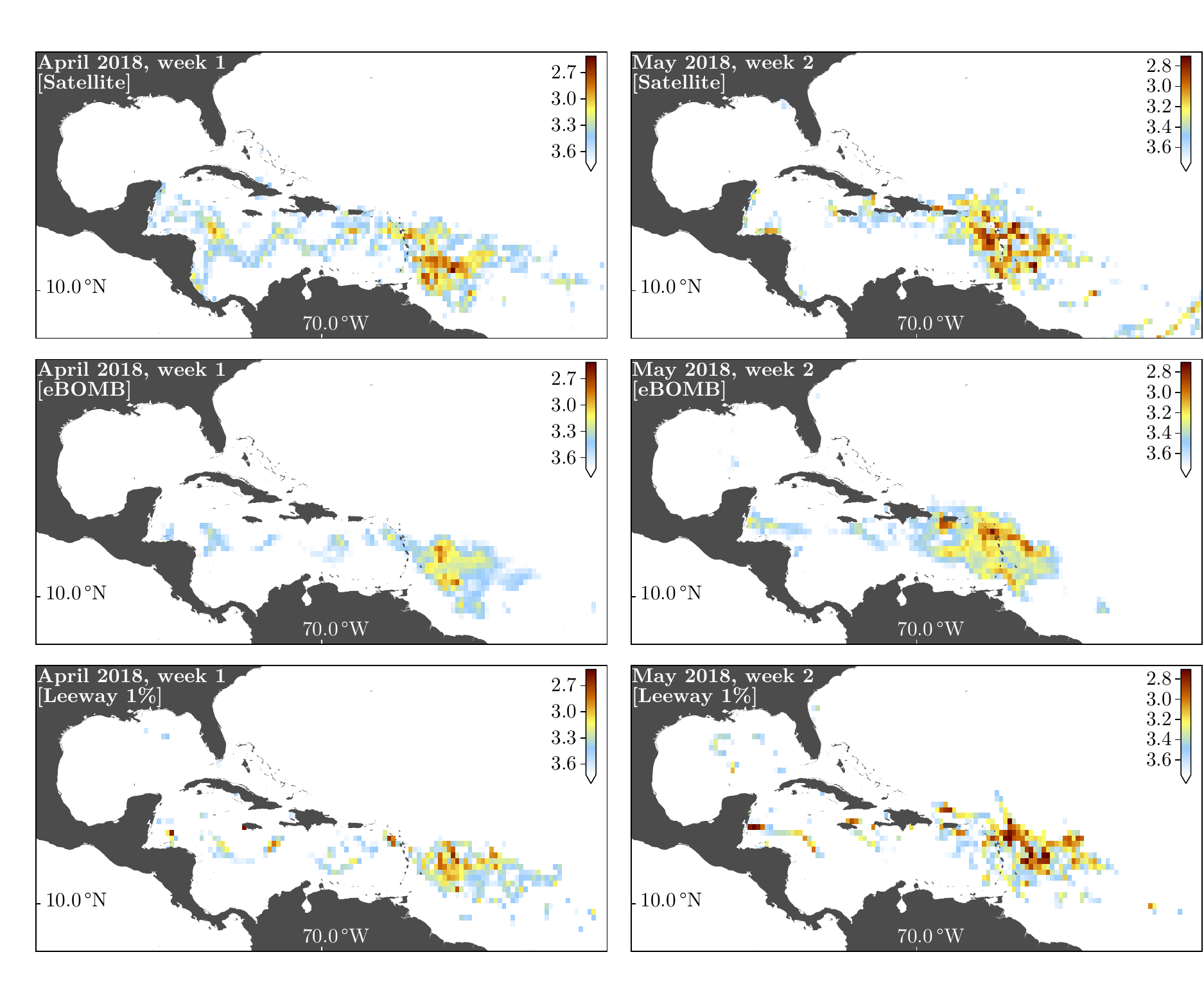}
    \caption{Comparison of the \ebomb{} (with biology turned off) and leeway models to satellite-derived \sarg{} coverage distributions for two different weeks in April 2018. The colorbar is as defined in Fig.\@~\ref{fig:initial-sarg}, except that the scales in each week are different to maintain clarity of comparison. }
    \label{fig:heatmap-comparison}
\end{figure}

\begin{figure}[t!]
    \centering
    \includegraphics[width = .75\textwidth]{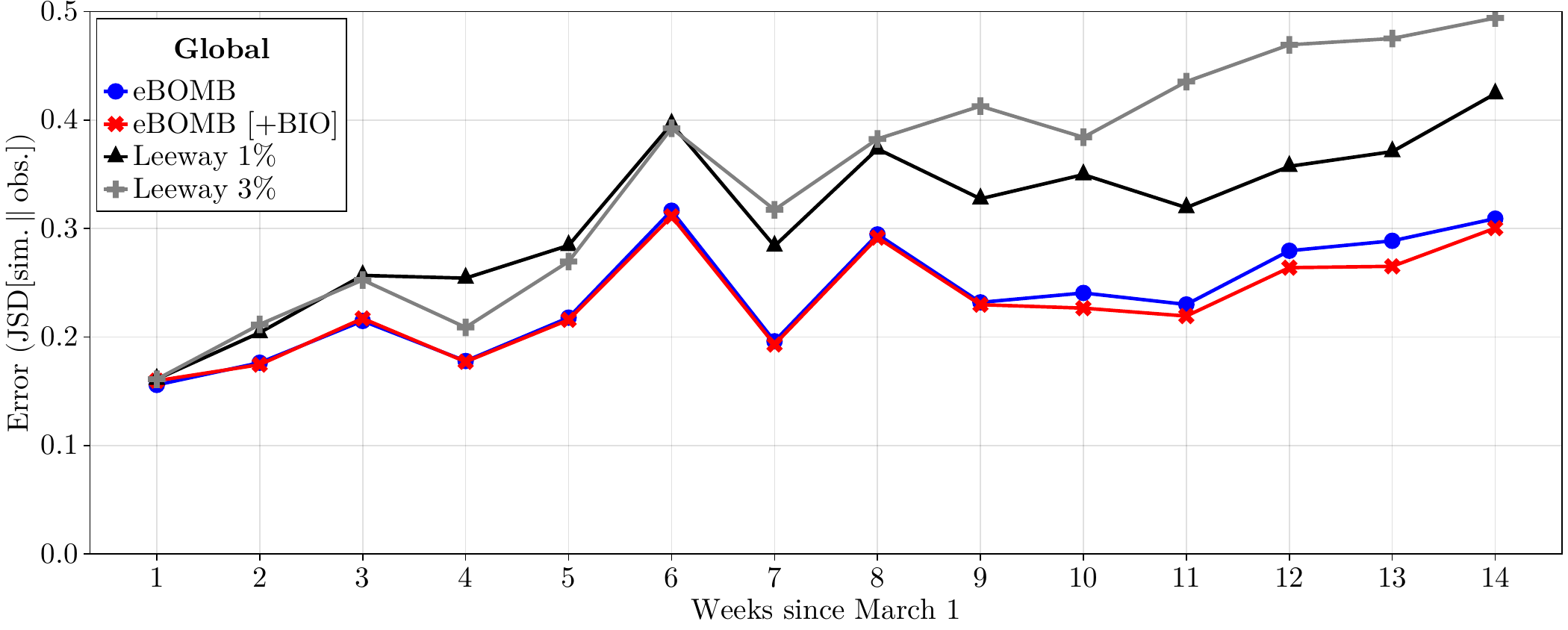}
    \caption{Weekly JSD similarity between observed \sarg{} coverage distribution and as predicted by the eBOMB model, with and without life cycles represented, and the leeway models, calculated in the region of the simualations.}
    \label{fig:timeseries-global}
\end{figure}

A qualitative analysis of the results for two representative weeks is illustrated in Fig.\@~\ref{fig:heatmap-comparison}, where the \ebomb{} model is configured to use only physics. We consider Leeway 1\% as it appears to be somewhat more accurate than Leeway 3\% \textcolor{black}{as per our more quantitative results shown in Fig.\@~\ref{fig:timeseries-global}}. The comparison indicates an overall enhancement achieved by the \ebomb{} model. Notice the wave-like pattern in the observed \emph{Sargassum} distribution during the first week of April (Fig.\@~\ref{fig:heatmap-comparison}, left column). This pattern is more accurately represented by eBOMB compared to Leeway 1\%. Similarly, the Leeway 1\% distribution near the Antilles Arc is shifted northwards relative to the observed distribution, while the eBOMB model captures its main location more precisely. The eBOMB model also shows better overall performance compared to Leeway 1\% in the second week of May (Fig.\@~\ref{fig:heatmap-comparison}, right column). Notably, Leeway 1\% predicts high concentrations of \emph{Sargassum} in the Gulf of Mexico, which the eBOMB model does not, aligning with observations. Additionally, Leeway 1\% infers higher concentrations in the western Caribbean Sea and east of the Antilles Arc. The eBOMB distributions generally align more closely with observed distributions.

A more rigorous quantitative comparison is achieved using the Jensen--Shannon Divergence (JSD) \cite{lin1991divergence}, a method for assessing the similarity between two probability distributions (see Materials and Methods for details). The JSD is inherently non-negative, with a value of 0 signifying identical distributions, and values greater than 0 indicating divergence between the distributions. We commence by calculating the JSD between the observed \emph{Sargassum} distribution and the predictions generated by the \ebomb{} model, both with and without biological factors, as well as Leeway 1\% and Leeway 3\%, across the entire simulation domain. This domain encompasses the Caribbean Sea, the Gulf of Mexico, and the North Atlantic from the equator to 32.5$^\circ$N and west of 40$^\circ$W. The results are presented adjusted for cloud coverage at weekly intervals in Fig.\@~\ref{fig:timeseries-global}. Two observations are particularly noteworthy. First and foremost, the \ebomb{} model markedly outperforms both Leeway 1\% and Leeway 3\% (with the former being more accurate than the latter), providing robust validation for our proposed model. Secondly, the inclusion of biological aspects enhances the performance of the eBOMB model relative to observations, albeit this enhancement is marginal. 

\begin{figure}[t!]
    \centering
    \includegraphics[width = .75\textwidth]{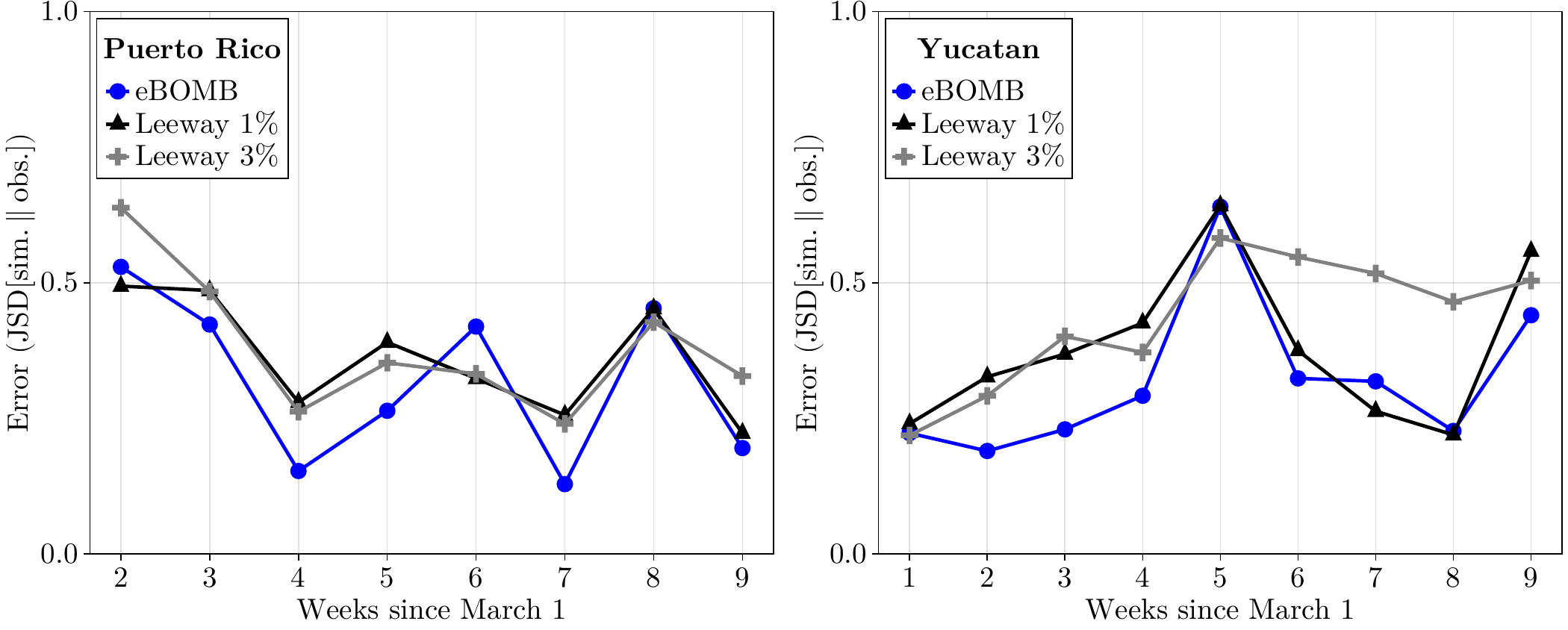}\\
    \includegraphics[width = .75\textwidth]{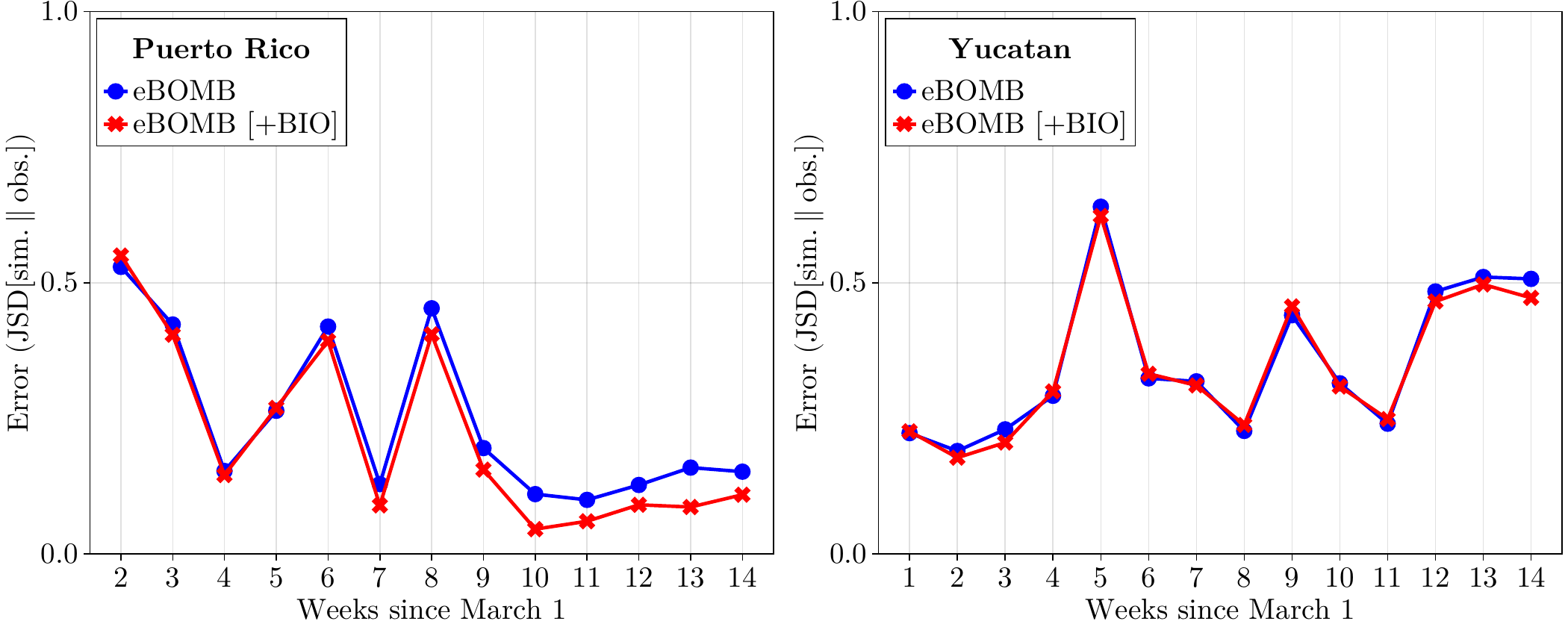}
    \caption{(top panels) As in Fig.\@~\ref{fig:timeseries-global}, but within two regions of interest, indicated with boxes in Fig.\@~\ref{fig:initial-sarg}. (bottom panels) As in the top panels, but restricted to the eBOMB model showing the effect of adding biology.}
    \label{fig:timeseries-local}
\end{figure}

We further test the local accuracy of each model in two regions of interest, highlighted with boxes in Fig.\@~\ref{fig:initial-sarg}.  One region encompases the Yucatan Peninsula, with its eastern coastline being notoriously affected by \sarg{} chocking.  The other region surrounds Puerto Rico Island, also reported to be affected by the accumulation of \sarg{}, especially in its sourthern coasts.  We again compute the JSD, renormalized in each of these regions \textcolor{black}{by the total probability they contain.}  The results are shown on a weekly basis in Fig.\@~\ref{fig:timeseries-local}.  Note the overall better performance of the eBOMB model with no biology compared to Leeway 1\% and Leeway 3\% (Fig.\@~\ref{fig:timeseries-local}, top panels).  In bottom panels of Fig.\@~\ref{fig:timeseries-local} we restrict the comparison to eBOMB model with and without life cycles represented.  As noted above, only a marginal overall improvement is provided by the activation of biology.  

It is important to acknowledge that the quality of the results presented is contingent upon several factors. Regarding physical parameters, the representations of ocean and wind velocities are far from perfect, particularly the ocean velocity, which, despite being a data synthesis rather than a simulation outcome, omits critical aspects of ocean circulation such as submesoscale motions. Conversely, it has been observed that the Stokes drift component of ocean velocity exerts a negligible effect; however, there remains a possibility that such drift is inaccurately represented by reanalysis systems. Additionally, the wind velocity data provided at 10-m above sea level from reanalysis systems does not accurately represent the near-surface wind field. In any event, should an accurate operational coupled ocean--wave--atmosphere model become available, its output could be utilized to enhance the performance of the eBOMB model. The effects of wave breaking, for instance, may be parameterized in eBOMB by simulating \sarg{} sinking in a manner analogous to the treatment of \sarg{} mortality. 

Concerning biological factors, it is postulated that biological effects may accumulate over time, thereby reducing discrepancies with observations. However, it is crucial to recognize that the nutrient and temperature data employed to inform the life cycles model equation \eqref{eq:life} \textcolor{black}{contain} uncertainty. Moreover, our selection of the life cycles model is admittedly rudimentary. Enhancements are anticipated through the adoption of a more sophisticated model, which can be seamlessly integrated into the eBOMB framework. 

Notwithstanding the aforementioned limitations, the eBOMB model, in its current configuration, surpasses the leeway model, which is further incapable of accurately depicting the observed aggregation of \sarg{} within mesoscale eddies \cite{Andrade-etal-22}. Mesoscale eddies serve as significant mechanisms for the long-range transport of ocean properties \cite{Gordon-86, Beal-etal-11, Wang-etal-16}, potentially contributing to the connectivity between distant regions of \sarg{} concentration. Below, we demonstrate the capability of eBOMB particles to represent concentrations within mesoscale eddies.

\subsection{Behavior near mesoscale eddies}

In \cite{Beron-Miron-20}, a rigorous result pertaining to the dynamics of eBOM particles in the proximity of mesoscale eddies is presented. This result is equally applicable to \emph{immortal} eBOMB particles, namely, those that are not influenced by physiological changes. An informal articulation of this result is presented below (see the Supplementary Materials for a rigorous statement and the corresponding proof).
\begin{theorem} \label{thm:vortices}
    In the presence of quiescent wind conditions, the cores of anticyclonic mesoscale eddies with coherent material boundaries function as finite-time local attractors for immortal eBOMB particles, whereas the cores of cyclonic eddies serve this role provided the particles exhibit sufficiently strong elastic connectivity. 
\end{theorem}
By a coherent material boundary, we denote a loop formed by water parcels that experience an identical cumulative material rotation relative to the material rotation of the surrounding water mass \cite{Haller-etal-16}. Numerical observations have demonstrated that each subset of these specialized material loops exhibits approximately uniform stretching over a finite-time interval \cite{Andrade-etal-20}. This attribute prevents them from experiencing breakaway filamentation \cite{Haller-Beron-13}. (See Supplementary Materials for a review of the notions of material coherence.) The property of immortal eBOMB particles articulated in Theorem \ref{thm:vortices} has significant implications for their long-range transport, and consequently for the transport of \emph{Sargassum} rafts, as illustrated in the Caribbean Sea by \cite{Andrade-etal-22}. This property is not shared by leeway particles, which, under the conditions specified in Theorem \ref{thm:vortices}, are unable to aggregate within mesoscale eddies, rendering these eddies ineffective as transport mechanisms.

\begin{figure}[t!]
    \centering
    \includegraphics[width = \textwidth]{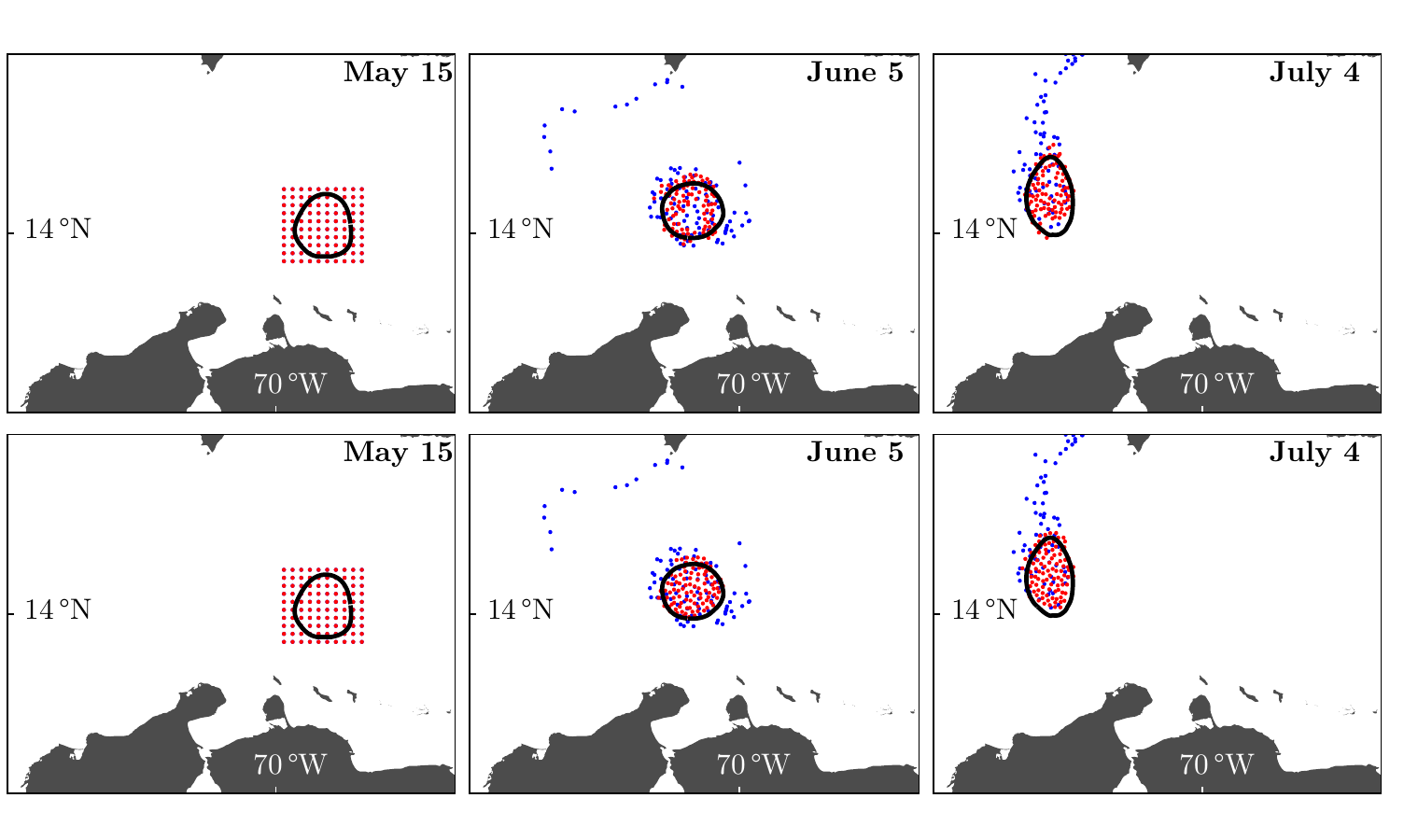}
    \caption{Depictions of the temporal progression of immortal eBOMB particles (red dots) and leeway particles (blue dots) initially encompassing an anticyclonic mesoscale eddy, extracted on 15/May/2018 in the eastern Caribbean Sea from a synthesis of altimetry, wind, and drifter ocean velocity data utilizing geodesic detection with a coherence horizon of 2 months. The immortal eBOMB particles in the upper row have lower stiffness compared to those in the lower row.}
    \label{fig:acy}
\end{figure}

\begin{figure}[t!]
    \centering
    \includegraphics[width = \textwidth]{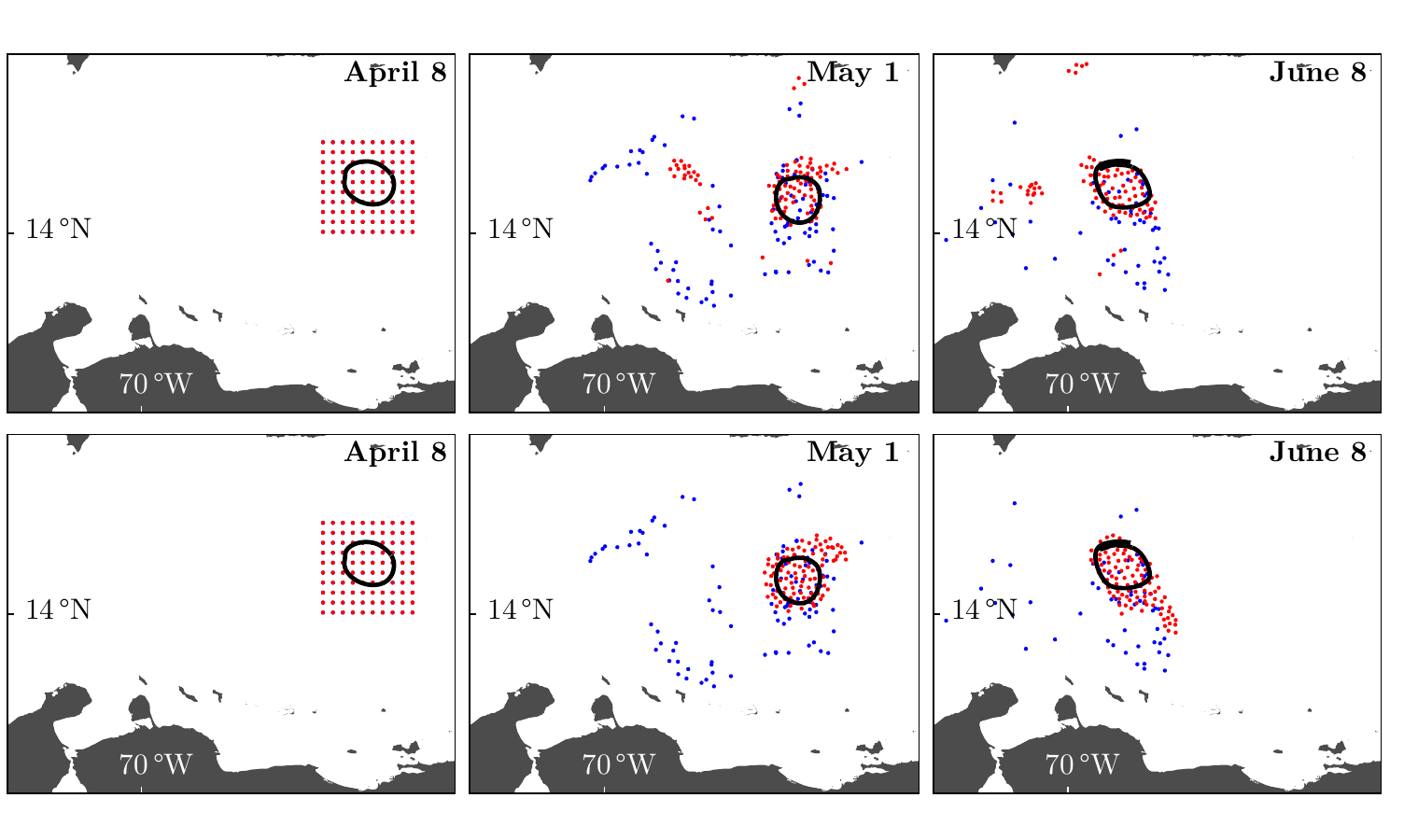}
    \caption{As in Fig.\@~\ref{fig:acy}, but for the case of a cyclonic eddy.}
    \label{fig:cyc}
\end{figure}

The validity of Theorem \ref{thm:vortices} is evaluated in Fig.\@~\ref{fig:acy} (resp., \ref{fig:cyc}) for the scenario of a mesoscale anticyclonic (resp., cyclonic) eddy. Both eddies were extracted from the same altimetry/wind/drifter synthesis of ocean velocity utilized in the aforementioned simulations. The extraction technique employed is geodesic eddy detection \cite{Haller-Beron-13}, ensuring that the extracted eddies possess material boundaries that exhibit uniform stretching over the extraction time interval. The red dots in each figure denote immortal eBOMB particles, while the blue dots represent leeway particles. The eBOMB particles in the top row of each panel have lower stiffness compared to those in the bottom row. Specifically, $A = 15.1\,\text{d}^{-2}$ (resp., $A = 200\,\text{d}^{-2}$) in the top (resp., bottom) row. Further, the number of nearest neighbor connections has been increased, and the natural length of the spring has been recomputed as described in Materials and Methods for this new initial distribution. The remaining eBOMB parameters are consistent with those used in the previous simulations, except that physiological transformations were considered. The leeway windage is set to $\alpha_{10} = 0.01$. Observe the tendency of immortal eBOMB particles to aggregate within the eddies. The expulsion of leeway particles is attributable to wind action, as observed during the simulation period (leeway particles with $\alpha_{10} = 0$ represent water particles, which are unable to traverse coherent material boundaries; thus, if initially inside, they would remain enclosed). The outcomes of the tests align with the predictions of Theorem \ref{thm:vortices}, despite being conducted beyond the strict domain of the theorem's applicability, which is noteworthy.

\section{Summary and Outlook}
\label{sec:discussion}

In this study, we have developed a mechanistic model for pelagic \emph{Sargassum} dynamics that integrates physical and biological processes. From a physical perspective, the model considers ocean currents and wind patterns to simulate \emph{Sargassum} movement influenced by inertial effects. This was achieved by extending a Maxey--Riley theory for floating finite-size particles to incorporate nonlinear forces to simulate interactions within and between \emph{Sargassum} rafts. Furthermore, the model includes a biological component by simulating \emph{Sargassum} reproduction through vegetative growth and fragmentation, providing a comprehensive view of its life cycle. The model was validated using satellite-derived data, demonstrating superior performance compared to the leeway approach to transport, which is the foundation of current \emph{Sargassum} modeling. Unlike our proposed model, the leeway model cannot simulate observed aggregation within mesoscale eddies, which provide a mechanism for long-range transport. Additionally, to encourage transparency and support broader scientific research, the model codes have been made publicly accessible, allowing the community to refine and apply the model in various research contexts, as well as in management and response efforts.

Future research endeavors will leverage the Maxey--Riley modeling framework to address critical questions regarding \emph{Sargassum} connectivity among the Sargasso Sea, the tropical North Atlantic, and the Intra-America Sea. Furthermore, we will examine the potential for riverine runoff in the Gulf of Guinea to initiate blooms, which, under the influence of ocean currents and winds modulated by inertial effects, could result in elevated \emph{Sargassum} concentrations throughout the tropical North Atlantic. This hypothesis necessitates validation through the application of suitable models and analytical tools. Notably, historical records indicate the presence of \emph{Sargassum} in the Gulf of Guinea in 2009, predating the first significant invasion of the Caribbean Sea by \emph{Sargassum} rafts \cite{Gloria-etal-16}, and preceding the extreme North Atlantic Oscillation (NAO) event hypothesized to have transported \emph{Sargassum} from the Sargasso Sea to the tropical Atlantic \cite{Johns-etal-20}. The model developed in this study is aptly suited for this purpose, and specialized methodologies from probabilistic dynamical systems \cite{VandenEijnden-06} can facilitate the conceptualization of connectivity \cite{Beron-etal-22-ADV, Bonner-etal-23}.

\section{Materials and Methods}
\label{sec:materials-and-methods}

\subsection{\sarg{} distributions from AFAI}
\label{sec:sarg-from-afai}

The Alternative Floating Algae Index (AFAI) is a metric derived from multiple spectral bands captured by the MODIS (Moderate-Resolution Imaging Spectroradiometer) instruments aboard NASA’s \emph{Aqua} and \emph{Terra} satellites, which can be utilized for the detection of floating algae, including \sarg{} \cite{hu2009novel, wang2016mapping}. AFAI and its extensions have been used to create \sarg{} coverage distributions with high precision and superior cloud rejection \cite{descloitres2021revisited, ody2019situ, hu2023map}. Here, we begin with AFAI data accumulated in 7-day increments for the year 2018 \cite{hu2015spectral, afai-data}. We then apply a simplified version of data processing pipeline of \cite{wang2016mapping} to produce weekly \sarg{} coverage distributions, mainly taking the distribution to be given by the difference between local AFAI values and the median background in a certain pixel window, subject to thresholds.

\begin{figure}[t!]
    \centering
    \includegraphics[width = \textwidth]{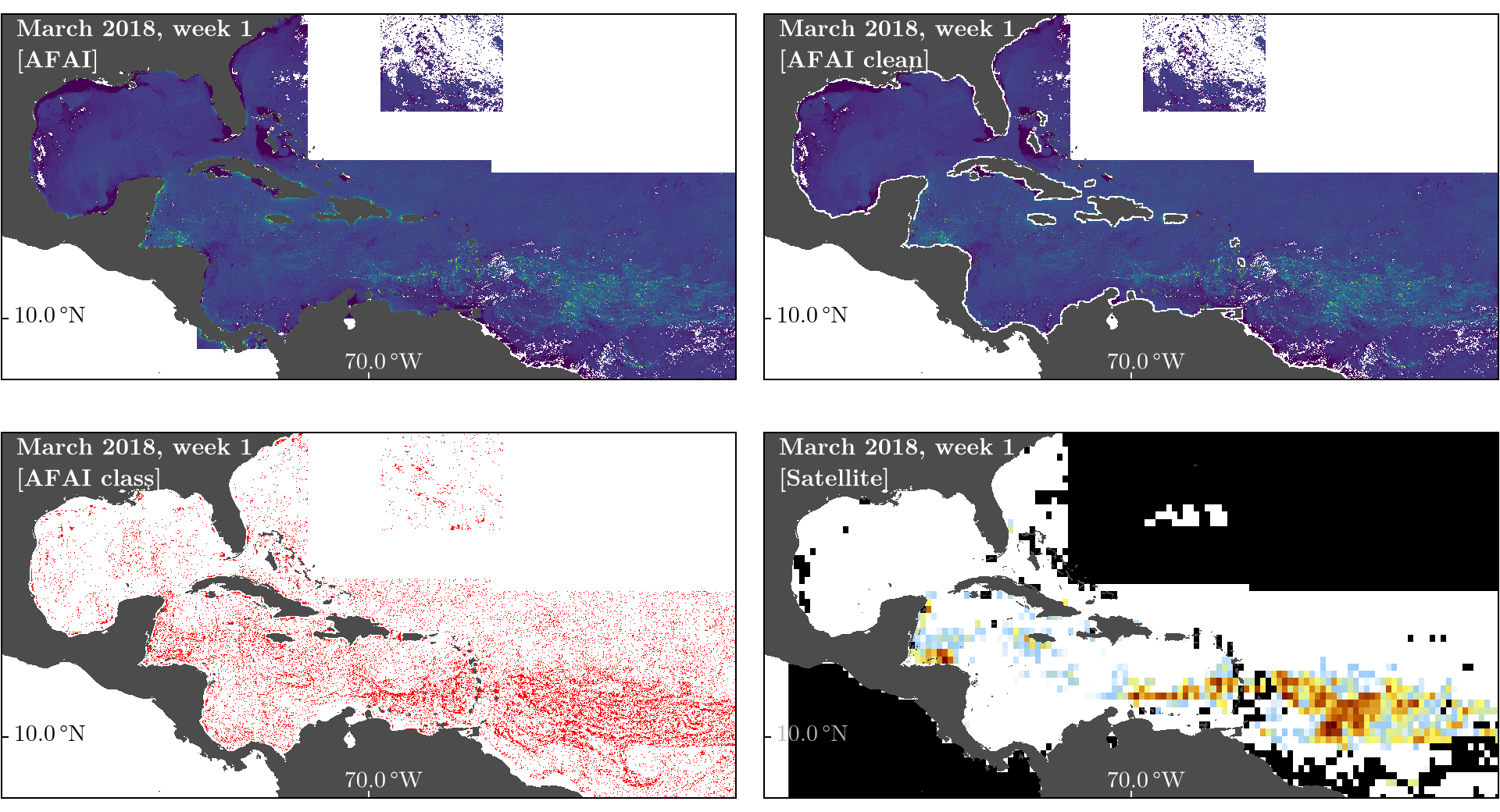}
    \caption{The main stages of the AFAI processing pipeline: 1) raw AFAI data (top left); 2) Pacific and coastal data removed (top right); 3) locations of \sarg{}-containing pixels in red (bottom left); and 4) the final thresholded and binned distribution, as in Fig.\@~\ref{fig:initial-sarg} (bottom right). Bins with significant missing data are colored in black. }
    \label{fig:afai-pipeline}
\end{figure}

Specifically, computation of a \sarg{} coverage distribution from raw AFAI data proceeds in three steps which are shown in Fig.\@~\ref{fig:afai-pipeline}. First, the data are cleaned by removing values in the Pacific and data within 20 pixels of a coastal region. Then, each pixel is classified according to whether it contains \sarg{}. To do this, the value of each pixel is decreased by the median value of all pixel's surrounding it within a window of size 51 pixels. If the resulting value is greater than $1.79 \times 10^{-4}$, it is considered to contain \sarg{}. Finally, the distribution is created by scaling the AFAI values according to a linear interpolation between $-8.77\times 10^{-4}$ and $4.41\times10^{-2}$ (recall that AFAI can take negative values), thresholding the top 85\% of data to reduce noise and binning at roughly $1/2^\circ$ resolution. Bins with greater than half of the data missing are considered to be cloud-covered. The data are normalized to a probability distribution such that the total coverage for each month is equal to $1$.

\subsection{Metric for model validation}
\label{sec:metric}

The Jensen--Shannon Divergence (JSD) \cite{lin1991divergence} is employed as the main metric to measure the difference between simulations and observations. Results obtained with other metrics, which exhibit similar patterns, are included in Supplementary Materials. Let $P_1$ and $P_2$ be two discrete probability distributions defined on a sample space $\mathcal{X}$. Let $Q = \frac{1}{2}(P_1 + P_2)$ be the mixture distribution of $P_1$ and $P_2$, i.e., the cumulative distribution function of $Q$ is equal to the sum of the cumulative distribution functions of $P_1$ and $P_2$, each weighted by $\frac{1}{2}$. Then we have
\begin{equation} \label{eq:jsd-defn}
    \jsd[P_1 \| P_2] = \frac{1}{2} \sum_{x \in \mathcal{X}}P_1(x) \log \left( \frac{P_1(x)}{Q(x)} \right) + \frac{1}{2} \sum_{x \in \mathcal{X}}P_2(x) \log \left( \frac{P_2(x)}{Q(x)} \right).
\end{equation}
We note that this is defined for one or both of the cases $P_1(x) = 0$ and $P_2(x) = 0$ by setting the term in the associated sum equal to zero. 

\subsection{Carrying flow representation}
\label{sec:determination-of-parameters}

The characterization of the Eulerian ocean velocity $\mathbf v_\text E$ is provided by an amalgamation of geostrophic flow, inferred from multisatellite altimetry data \cite{LeTraon-etal-98}, and Ekman drift, induced by wind reanalysis \cite{Dee-etal-11}. This synthesis is calibrated to align with the velocities of drifters from the NOAA Global Drifter Program \cite{Lumpkin-Pazos-07}, which are drogued at 15-m depth. The resultant altimetry/wind/drifter synthesis is available for distribution through \href{ftp://ftp.aoml.noaa.gov/phod/pub/lumpkin/decomp}{ftp://ftp.aoml.noaa.gov/phod/pub/lumpkin/decomp}.  \textcolor{black}{The product has been validated in experiments involving small object tracking in the Atlantic \cite{Olascoaga-etal-20, Miron-etal-20-GRL}.} The Stokes drift velocity $\mathbf{v}_\text{S}$ and the wind velocity at a 10-meter elevation $\mathbf{v}_{10}$ are produced by the ECMWF Reanalysis v5 (ERA5), accessible via \href{https://www.ecmwf.int/en/forecasts/dataset/ecmwf-reanalysis-v5}{https://www.ecmwf.int/en/forecasts/dataset/ecmwf-reanalysis-v5}. In our simulations we have taken $\mathbf{v}_{10}$ to be representation of the near-surface wind velocity $\mathbf v_\text W$.

\subsection{Determination of parameters via optimization}
\label{sec:determination-of-parameters}

In order to determine the free parameters of the \ebomb{} model, we optimize simulations by treating the AFAI-derived \sarg{} distributions as a target. We initialize our clumps to match the \sarg{} coverage distribution in the first week of March 2018. This is done by dividing the nonzero coverage values into 10 logarithmically spaced levels labeled $i = 1, 2, \dotsc, 10$ so that each longitude--latitude bin has an $i$ associated to it by that level. Then, we place $i$ clumps uniformly at random in each bin, resulting in 2710 initial clumps. A simulation was then run for a time up to and including the second week of May, 2018, recording the location of each clump every 0.1 d. Note that our time-series comparisons, e.g., Fig.\@~\ref{fig:timeseries-global} extend beyond this record. For the time between 7 March 2018 and 14 March 2018, every clump position was pooled and binned to the same resolution as the AFAI-derived \sarg{} distribution during this week. The simulated and satellite distributions were both normalized and the JSD was computed using Eq.\@~\ref{eq:jsd-defn}, excluding bins that were covered by clouds. This process was repeated for each week in the full integration time span, and the total loss for the simulation is taken to be the sum of all weekly JSD values. In this way, we obtain a single quantity which, when small, indicates that the simulation is accurate relative to the observations in a cumulative sense. Then, for a given set of parameters to optimize, we use the cumulative JSD value as the target to minimize. 

To obtain our final values in Table~\ref{table:parameters}, we optimized in two stages. In the first stage, we turn off biological growth and death and optimize the physics parameters $\tau, \delta, \sigma$ and $A$. In the second stage we fix the physics parameters, turn on biological growth and death, and optimize $\mu_\text{max}, m, k_\text N, S_\text{min}$ and $S_\text{max}$. This staged optimization ensures that the core physics model is accurate without relying too heavily on biological factors. We therefore have two derivative-free black-box optimization problems with relatively expensive function evaluations. To solve this, we use the NOMAD implementation of the mesh adaptive direct search algorithm \cite{nomad4paper}, which interfaces with the SciML optimization packages. The NOMAD solver accepts box constraints; we set these according to our confidence in the value of each parameter. The buoyancy was measured in \cite{Olascoaga-etal-23} to lie in the range $1.00$ to $1.49$. The Maxey--Riley theory we apply requires the inertial response time to be small (see Supplementary Material); it is sufficient to take $\tau f \lesssim 1$ to obtain an upper bound and the lower bound is a factor of 10 below this. Similarly, we take the Stokes drift coefficient in the range 0.5 to 1.5. Other optimized parameters were constrained within approximately a factor of 10 on either size of heuristic starting values. Each stage of the optimization was run with a hard time limit of one hour on a 2022 MacBook Pro (Apple M2 chip, 16 GM RAM). We note that certain biological parameters in our study exhibit values considerably lower than those reported in the literature \cite[e.g.,][]{Jouanno-etal-23}. This discrepancy can be attributed to the incorporation of biological effects as a supplementary module to the mechanistic framework within our model, thereby introducing expected variations. Furthermore, our biological model is implemented in the context of discrete clumps, wherein the concept of fractional growth is less rigorously defined. We anticipate that the breadth and precision of the growth and death effects can be substantially enhanced in subsequent research, particularly to facilitate the direct inclusion of empirical measurements.

\subsection{Complete model specification}
\label{sec:full-model-specs}

Tables~\ref{table:fields} and \ref{table:parameters} summarize all of the datasets and parameters required to fully define our model and Algorithm~\namedref{alg:ebomb}{eBOMB} provides the pseudo-code. We comment briefly on an implementation detail regarding the life cycle portion of Algorithm~\namedref{alg:ebomb}{eBOMB}. The SciML ecosystem provides a \emph{callback} interface where the integration status can be conditionally modified during or after each step in an efficient and safe manner. This allows us to log the clump positions at regular time steps $(t_i, t_i + h)$ while the underlying integration checks for growth and death (due to either biology or beaching) at the end of whichever step size the integrator has chosen. Hence, we can still take advantage of fast adaptive integration algorithms such as the method of \cite{tsitouras2011runge}. Further, using these callbacks, Algorithm~\namedref{alg:ebomb}{eBOMB} can be implemented such that the length of $c$, that is, the length of the vector of clump positions, is dynamically modified whereby at given time it only contains the positions of living clumps. This can provide an increase in speed at the expense of introducing another variable whose length is always equal to the length of $c$ that keeps track of the labels of the clumps by vector position.

\begin{table}
    \centering
    \begin{tabular}{llll}
        \hline\hline
        \textbf{Name}           & \textbf{Symbol}   & \textbf{Units}        & \textbf{Source}           \\  
        Eulerian velocity       & $v_\text E$       & km\,d$^{-1}$          & \cite{lumpkin-data}       \\
        Stokes drift            & $v_\text S$       & km\,d$^{-1}$          & \cite{wind-stokes-data}   \\
        Wind velocity           & $v_\text W$       & km\,d$^{-1}$          & \cite{wind-stokes-data}   \\
        Temperature             & $T$               & $^{\circ}$C           & \cite{temp-data}          \\
        Nitrogen content        & $N$               & mmol\,m$^{-3}$        & \cite{nutrient-data}      \\
        Landmass location       & land              & None                  & \cite{land-data}          \\
        \hline
    \end{tabular}
    \caption{Model datasets. Water and wind velocities are shared by \ebomb{} and Leeway.}
    \label{table:fields}
\end{table}

\begin{table}
    \centering
    \begin{tabular}{lllll}
        \hline\hline
        \textbf{Parameter name}             & \textbf{Symbol}       & \textbf{Value}              & \textbf{Units}     & \textbf{Source}            \\   
        Windage parameter                   & $\alpha$              & $3.37\times10^{-3}$         & None               & Derived                    \\
        Inertial response time              & $\tau$                & $0.0103$                    & d                  & Optimization               \\
        Buoyancy                            & $\delta$              & 1.14                        & None                  & Optimization               \\
        Clump radius                        & $a$                   & $6.4\times10^{-5}$          & km                 & Derived$^\star$            \\
        Spherical factor                    & $R$                   & 0.823                       & None                 & Derived                    \\
        Coriolis parameter                  & $f$                   & 2.18213                     & d                  & Derived                    \\
        Stokes drift coefficient            & $\sigma$              & 1.2                         & None                  & Optimization               \\
        Stiffness amplitude                 & $A$                   & 15.1                        & $\text{d}^{-2}$    & Optimization               \\
        Stiffness cutoff scale              & $\Delta$              & 0.2                         & km                 & Heuristic                  \\
        Spring natural length               & $L$                   & Variable                    & km                 & Heuristic                  \\
        Maximum growth rate                 & $\mu_{\text{max}}$    & 0.00541                     & $\text{d}^{-1}$    & Optimization               \\
        Mortality rate                      & $m$                   & 0.00402                     & $\text{d}^{-1}$    & Optimization               \\
        Nitrogen uptake half saturation     & $k_{\text{N}}$        & 0.000129                    & mmol/m$^3$         & Optimization               \\
        Temperature growth minimum          & $T_{\text{min}}$      & 10.0                        & $^{\circ}\text{C}$ & \cite{Jouanno-etal-21b}    \\
        Temperature growth maximum          & $T_{\text{max}}$      & 40.0                        & $^{\circ}\text{C}$ & \cite{Jouanno-etal-21b}    \\
        Clump amount minimum                & $S_{\text{min}}$      & $-0.00482$                  & None                & Optimization               \\
        Clump amount maximum                & $S_{\text{max}}$      & 0.001                       & None                 & Optimization               \\
        \hline
    \end{tabular}
    \caption{Parameters of the \ebomb{} model. The designation ``Derived$^\star$'' signifies that $a$ constitutes a more fundamental parameter in comparison to $\tau$; nevertheless, $\tau$ is optimized due to its more explicit occurrence in the eBOMB model equations.}
    \label{table:parameters}
\end{table}

\begin{algorithm} 
\label{alg:ebomb}
\caption{\ebomb{} model for \sarg{} clump trajectories with life cycles}
\begin{algorithmic}[1]
\Require Datasets from Table~\ref{table:fields} and constants from Table~\ref{table:parameters}.
\Require Integration time span $(t_1, t_2)$.
\Require Integration algorithm with time step $h$. \Comment{Such as RK4.}
\Require The maximum number of allowed clumps $N_{\text{max}}$.
\Require Clump initial positions $c = (x_1, y_1, x_2, y_2, \dots, x_{N_{\text{max}}}, y_{N_{\text{max}}})$. 
\Require Clump initial connections (\neigh{$i$, $t_1$})$_{1\leq i \leq N_{\text{max}}}$.
\Require Clump initial amounts $S = (0)_{1\leq i \leq N_{\text{max}}}$. \Comment{Or a distribution of $S$ values.}
\Require Clump initial life status $\ell = (\ell_1, \ell_2, \ell_3, \dots \ell_{N_{\text{max}}})$ where $\ell_i \in \{1, 0 \}$ according to whether clump $i$ is alive.
\State Compute $L$ using $c$ and the procedure described in \nameref{sec:model-overview} with $K = 5$.
\State Set $t \gets t_1$.
\State Set $N \gets \text{sum}(\ell)$. \Comment{This is the total number of clumps that have ever existed.}
\While{$t < t_2$}
    \State Integrate $\dot{c} = g(t, c)$ with $g$ implied by Eq.\@~\eqref{eq:MR-high-level-def} for one time step $h$.
    \State Set $c \gets c(t + h)$ and $t \gets t + h$.
    \State Set $\ell_i \gets 0$ for all clumps for which $(x_i, y_i) \in \text{land}$. 
    \For{$\{i : \ell_i = 1 \}$}
        \State Update $S_i$ using Eq.\@~\eqref{eq:S-clump-amount-def} evaluated at $(t, x_i, y_i)$.
        \If{$S_i < S_\text{min}$}
            \State Set $\ell_i \gets 0$.
        \ElsIf{$S_i > S_\text{max}$}
            \State Set $S_i \gets 0$. \Comment{Refresh this ``parent'' clump.}
            \State Set $N \gets N + 1$.
            \State Set $\ell_{N} \gets 1$.
            \State Generate $\theta \in [0, 2\pi)$ uniformly at random. 
            \State Set $x_{N} \gets x_i + L \cos \theta$ and $y_{N} \gets y_i + L \sin \theta$.
        \EndIf
    \EndFor

    \State Form new connections (\neigh{$i$, $t$} : $\ell_i = 1$) among living clumps.
\EndWhile
\State Return $c$
\end{algorithmic}
\end{algorithm}

\subsection{Code and data availability}

\textcolor{black}{
All code utilized for the simulation, optimization, and visualization of the \ebomb{} model is disseminated freely in an open-source manner under the MIT License via the GitHub repository \href{hhttps://github.com/70Gage70/Sargassum.jl}{https://github.com/70Gage70/Sargassum.jl}. The main tool for end users is the Julia package \texttt{Sargassum.jl}, distributed initially as version \texttt{0.1.2}. This is a meta-package collecting all of the required sub-packages in one place, including functionality for simulations, interfacing, and AFAI data processing. In particular, we provide an interface that allows our software to be used with zero coding in the form of a reactive Pluto.jl notebook \cite{pluto2024}. All packages were developed using \texttt{Julia 1.10.2} \cite{bezanson2017julia}. All raw data corresponding to each field in Table~\ref{table:fields} employed in this study are readily accessible for download through the provided interfaces. The integration of ordinary differential equations and black-box optimization interfaces is facilitated by the SciML ecosystem \cite{rackauckas2017differentialequations}. The data visualizations presented in this paper were generated using the Makie ecosystem \cite{DanischKrumbiegel2021}.
} 

\subsection*{Acknowledgements}

We extend our gratitude to Nathan Putman for the benefit of many useful discussions regarding \emph{Sargassum} phenomenology. This research was funded by the National Science Foundation (NSF) under grant number OCE2148499.

\section*{Supplementary Materials}

\appendix
\numberwithin{equation}{section}

\section{Parameters of the eBOMB model}
\label{sec:additional-maxey-riley-equations}

Here we provide the equations required to relate $\alpha$, $\tau$, and $R$ to the more fundamental quantities $a$ and $\delta$. For $\delta \ge 1 \sim O(1)$, we define
\begin{subequations} \label{eq:MR-geometry}
\begin{align}
\psi(\delta) &= \sqrt[3]{\mathrm{i} \sqrt{1-\left(\frac{2}{\delta }-1\right)^2}+\frac{2}{\delta }-1}, \\ 
\Phi(\delta) &= \frac{1}{2} \left(\mathrm{i} \sqrt{3}\right) \left(\frac{1}{\psi (\delta )}-\psi (\delta )\right)-\frac{1}{2} \left(\psi (\delta )+\frac{1}{\psi (\delta )}\right)+1, \\ 
\Psi(\delta) &=  \pi^{-1} \cos ^{-1}(1-\Phi (\delta )) -  \pi^{-1} (1-\Phi (\delta )) \sqrt{1-(1-\Phi (\delta ))^2}
\end{align}
\end{subequations}
where $\mathrm{i} = \sqrt{-1}$. Then,
\begin{subequations} \label{eq:MR-atRf}
\begin{align}
\alpha(\delta) &= \frac{\gamma \Psi}{1 - (1 - \gamma) \Psi}, \\
\tau(\delta, a) &= \frac{1 - \tfrac{1}{6} \Phi}{(1 - (1 - \gamma) \Psi) \delta^4} \frac{a^2 \rho}{3 \mu} , \\ 
R(\delta) &= \frac{1 - \tfrac{1}{2} \Phi}{1 - \tfrac{1}{6} \Phi},
\end{align}
\end{subequations}
where $\rho$ is the surface density of the water, $\mu$ is the water dynamic viscosity, and $\gamma$ is the air-to-water viscosity ratio. Typical values for these parameters are $\rho = 1.027 \times10^{12} \, \text{kg} \, \text{km}^{-3}$, $\mu = 8.873 \times 10^{4} \, \text{kg} \, \text{km}^{-1}\, \text{d}^{-1}$, and $\gamma = 1.7527 \times 10^{-2}$. 

\section{Derivation of the eBOMB model}
\label{sec:sphere}

Equation (2) constitutes an approximation, applicable for sufficiently small clumps, to the full Maxey--Riley equation governing a network of elastically interacting clumps on the ocean surface. As Newton's second law, this equation is \emph{second order} in clump position, thereby representing a nonautonomous $4N$-dimensional coupled dynamical system. The $2N$-dimensional system Eq.\@~(2) dictates the evolution on the so-called slow manifold, which asymptotically attracts all solutions of the $4N$-dimensional system. This provides support to Eq.\@~(2), along with advantages related to computational efficiency and the lack of necessity to specify the initial velocity of the clumps, which is typically unavailable.

Specifically, the full Maxey--Riley law for elastically interacting floating finite-size particles reads
\begin{equation}
     \underbrace{\dot{\mathbf v}_i + (f + \tau_\odot \hat{\mathbf x}\cdot\mathbf v_i)\mathbf v_i^\perp}_{(1)}  =  \underbrace{R\left.\left(\partial_t\mathbf v + \nabla_{\mathbf v}\mathbf v + f \mathbf v^\perp\right)\right\vert_i}_{(2)} +  \underbrace{\tfrac{1}{3}R\omega(\mathbf v\vert_i^\perp - \mathbf v_i^\perp)}_{(3)} +  \underbrace{\frac{\mathbf u\vert_i - \mathbf v_i}{\tau}}_{(4)} + \underbrace{\mathbf F_i}_{(5)},
    \label{eq:full}
\end{equation}
$i = 1, 2, \dotsc, N$.

System \eqref{eq:full} of second-order ordinary differential equations (with respect to the clump position $\mathbf{x}_i$) was presented in \cite{Beron-Miron-20} assuming Cartesian geometry. Additionally, a different perspective on the beads within the elastic network was offered, employing Hooke's law to characterize the elastic forces connecting the beads. This differentiates the eBOM model from the physics module found in the eBOMB model.

The interpretation of each term in Eq.\@~\eqref{eq:full} is as follows: term (1) denotes the (absolute) acceleration of clump $i$; term (2) delineates the flow force (per unit mass) and added mass effects on clump $i$; term (3) characterizes the lift force attributable to flow shear; term (4) corresponds to the drag force arising from fluid viscosity; and term (5) represents the elastic force, which, unlike in \cite{Beron-Miron-20}, incorporates a position-dependent stiffness, in contrast to the constant stiffness described by Hooke's law. Terms (2)--(4) are derived from the vertical integration of the fluid mechanics Maxey--Riley equation for a particle situated at the interface between the ocean and the atmosphere \cite{Beron-etal-19-PoF}.

\subsection{Covariant derivative}

We recall that the term 
\begin{equation}
    \partial_t \mathbf{v} + \nabla_{\mathbf{v}} \mathbf{v}
\end{equation}
is the \emph{covariant derivative of $\mathbf v$ along a trajectory $\mathbf x(t)$ generated by $\mathbf v$, that is, a solution to $\dot{\mathbf x} = \smash{\sqrt{\mathsf{m}^{-1}}}\,\mathbf v$}. This is represented as $\nabla_{\dot{\mathbf x}} \mathbf{v}$ or $\frac{D\mathbf v}{dt}$, which evaluates, using the chain rule,
\begin{equation}
    \nabla_{\dot{\mathbf x}} \mathbf{v} = \frac{D\mathbf v}{dt} = \partial_t \mathbf{v} + \nabla_{\mathbf{v}} \mathbf{v},
    \label{eq:Dvdt}
\end{equation}
where
\begin{equation}
    \nabla_{\mathbf v}\mathbf v = \sqrt{\mathsf m}\left(\dot x^j\partial_j\dot{\mathbf x} + \mathbf{\Gamma}_{jk} \dot x^j\dot x^k\right)
\end{equation}
is the \emph{covariant derivative of $\mathbf v$ in the direction of itself}. Here, $\smash{\mathbf{\Gamma}_{jk}} = (\smash{\Gamma^1_{jk}}, \smash{\Gamma^2_{jk}})$, where, in general,
\begin{equation}
    \Gamma _{ij}^{k}(\mathbf{x}) := \tfrac{1}{2} m^{kl}\left(\partial_jm_{il} + \partial_im_{jl} - \partial_lm_{ij}\right),
\end{equation}
where $\partial_i$ is short for $\partial/\partial x^i$, are the Christoffel symbols (of second kind), which establish the (Levi--Civita) connection on a $n$-dimensional Riemannian manifold $M$ coordinatized by $\mathbf x = (x^1, x^2, \dotsc, x^n)$ with metric representation given by the matrix $\mathsf m = (m_{ij})$ (the $(i,j)$th entry of $\mathsf m^{-1}$ is denoted $m^{ij}$). In essence, these emerge due to the fact that both the components of a vector on $M$ and the basis vectors themselves are functions of $x$.  On the sphere with rescaled geographic coordinates $(x,y)$, the only nonzero Christoffel symbols are 
\begin{equation}
    \Gamma_{11}^{2} = \gamma_\odot ^{2}\tau_\odot,\quad \Gamma_{12}^{1} = \Gamma_{21}^{1} = -\tau_\odot.
    \label{eq:G}
\end{equation}
With this in mind, one finds
\begin{equation}
    \nabla_{\mathbf v}\mathbf v = \gamma^{-1}v_x\partial_x\mathbf v + v_y\partial_y\mathbf v + \tau_\odot v_x\mathbf v^\perp.
\end{equation}
In the case of Cartesian geometry, where $\mathsf m$ is the identity matrix, the expression $\frac{D\mathbf v}{dt}$ in \eqref{eq:Dvdt} transforms into $\frac{D\mathbf v}{Dt} = \partial_t\mathbf v + (\mathbf v\cdot\nabla) \mathbf v$, which is the commonly known \emph{material derivative of $\mathbf v$ along its trajectory $\mathbf x(t)$}.

For details on covariant differentiation, readers are directed to Chapter 2.7 of \cite{Abraham-Marsden-87}.

\subsection{Slow manifold reduction}

For small clumps, i.e., with small radius $a$, the inertial response time ($\tau \propto a^2$) is short. We formally write this as $\tau = O(\mathrm{St})$ where $\mathrm{St} = a^2 \rho \mathcal{U}_\text{slip} / \mathcal{L} \mu > 0$ small is the Stokes number. Here, $\mathcal L$ is a characteristic length scale and $\mathcal{U}_\text{slip}$ is the magnitude of the slip velocity, i.e., that of the inertial particle velocity relative to that of the carrying flow. That $\tau = O(\mathrm{St})$ has an important consequence: Eq.\@~\eqref{eq:full} represents a singular perturbation problem involving slow, $\mathbf{x}_i$, and fast, $\mathbf{v}_i = \smash{\sqrt{\mathsf{m}^{-1}}}\,\dot{\mathbf x}_i$, variables. This readily follows by rewriting Eq.\@~\eqref{eq:full} as a system of first-order ordinary differential equations in $(\mathbf{x}_i, \mathbf{v}_i)$, i.e., a nonautonomous $4N$-dimensional dynamical system, which reveals that while $\mathbf{x}_i$ changes at $O(1)$ speed, $\mathbf{v}_i$ does it at $O(\mathrm{St}^{-1})$ speed. 

The aforementioned allows for the use of geometric singular perturbation theory (GSPT) as developed by Fenichel \cite{Fenichel-79, Jones-95} and extended to nonautonomous systems \cite{Haller-Sapsis-08}, on Eq.\@~\eqref{eq:full} to frame its \emph{slow manifold}.  That is, the normally hyperbolic $(2N + 1)$-dimensional manifold
\begin{equation}
    \big\{(\mathbf x_i, \mathbf v_i, t) : \mathbf v_i = \mathbf u(\mathbf x_i,t) + \tau \mathbf u_\tau(\mathbf x_i,t) + \tau \mathbf F_i + O(\mathrm{St}^2),\, i = 1, 2, \dotsc, N\big\}
    \label{eq:slow}
\end{equation}
of the $(4N + 1)$-dimensional phase space $(\mathbf x_i, \mathbf v_i, t)$, $i = 1, 2, \dotsc, N$, which attracts all solutions of Eq.\@~\eqref{eq:full} exponentially fast. The GSTP method, as detailed by \cite{Beron-etal-19-PoF}, was employed to derive a reduced equation governing the dynamics on the slow manifold of the BOM model. Further elucidation of the BOM model's slow manifold can be found in \cite{Olascoaga-etal-20}, with additional insights provided in the comprehensive review on inertial ocean dynamics by \cite{Beron-21-ND}.   

In \cite{Beron-Miron-20}, GSPT was utilized on Eq.\@~\eqref{eq:full} under the assumption of Cartesian geometry, though specifics were omitted. Although these specifics are available in the aforementioned papers, we find it instructive to outline at least one step that leads us to the conclusion that the motion on the slow manifold Eq.\@~\eqref{eq:slow} is governed by the eBOMB equation, Eq.\@~(2). 

Thus, we make $\varphi = t$ and write Eq.\@~\eqref{eq:full} in system form as
\begin{equation}
    \left\{
    \begin{aligned}
        \dot{\mathbf x}_i &= \sqrt{\mathsf{m}^{-1}}\,\mathbf v_i\\
        \dot{\mathbf v}_i &= - (f + \tau_\odot \hat{\mathbf x}\cdot\mathbf v_i)\mathbf v_i^\perp + R\left.\left(\partial_t\mathbf v + \nabla_{\mathbf v}\mathbf v + f \mathbf v^\perp\right)\right\vert_i + \tfrac{1}{3}R\omega(\mathbf v\vert_i^\perp - \mathbf v_i^\perp) + \frac{\mathbf u\vert_i - \mathbf v_i}{\tau} + \mathbf F_i,\\
        \dot \varphi &= t.
    \end{aligned}
    \right.
    \label{eq:sys}
\end{equation}
Substituting $\mathbf v_i = \mathbf u\vert_i + \tau \mathbf u_\tau\vert_i + O(\mathrm{St}^2)$ into Eq.\@~\eqref{eq:sys} and matching the terms with powers of $\tau = O(\mathrm{St})$, we obtain
\begin{equation} 
    \sqrt{\mathsf{m}}\,\dot{\mathbf{x}}_i = \mathbf{u}\vert_i + \tau\left.\left(R(\partial_t\mathbf v + \nabla_{\mathbf v}\mathbf v) + R \left( f + \tfrac{1}{3}\omega \right)\mathbf{v}^\perp - \dot{\mathbf{u}} - \left(f + \tau_\odot u_x + \tfrac{1}{3}R \omega \right) \mathbf{u}^\perp\right)\right\vert_i  +  \tau \mathbf{F}_i
\end{equation}
with an $O(\mathrm{St}^2)$ error.  Equation (2) is derived by writing
\begin{equation} 
    \dot{\mathbf{u}} = \nabla_{\dot{\mathbf x}}\mathbf u = \frac{D\mathbf u}{dt} = \partial_t\mathbf u + \nabla_{\mathbf u}\mathbf u, 
\end{equation}
\emph{which holds true by enforcing the closure $\dot{\mathbf x} = \smash{\sqrt{\mathsf{m}^{-1}}}\,\mathbf u$, which produces the trajectory $\mathbf x(t)$.} This step was implicitly made in \cite{Beron-etal-19-PoF} and subsequent papers, and we have explicitly stated it here.

\subsection{Comments on earlier formulations in spherical coordinates}

We have identified three inaccuracies in \cite{Beron-etal-19-PoF} that we take the chance to rectify: 1) the initial equality in Eq.\@~(A5) should be omitted; 2) the term $\dot{x}$ in Eq.\@~(A8) must be scaled by $\sqrt{\mathsf m}$; and 3) the term $\tau_\odot v^1$ preceding $\frac{1}{3}R\omega$ in Eq.\@~(A8) should be replaced with $2\tau_\odot u^1$ (the discrepancy in the total derivative definition with respect that used in this paper results in the factor of 2). The first error was propagated to \cite{Beron-21-ND}, as can be seen in Eq.\@~(39) in that paper.  The third error was also propagated to \cite{Miron-etal-20-GRL}; however, in that study, the complete BOM equation was employed, as opposed to its reduced form applicable asymptotically on the slow manifold, thereby not affecting the results.

\section{Rigorous formulation of Theorem 1 and proof}
\label{sec:proof-of-thorem}

To rigorously state the result on the role of mesoscale eddies (vortices) as attractors for eBOMB particles as outlined in Thm.\@~1, we follow \cite{Beron-Miron-20}, noting some differences, addressing certain omissions and minor errors, and clarifying the assumptions. The approach used in \cite{Beron-Miron-20} is based on the procedure described in \cite{Haller-etal-16}.

\subsection{Preparation}

We begin by reviewing two independent notions of material coherence.

\begin{definition}[Flow map]
    Let $F_{t_0}^t$ be the \textbf{flow map} that takes the positions $\mathbf x_0$ of the fluid particles at time $t_0$ to their new positions $\mathbf x$ at a later time $t$, obtained by solving
    \begin{equation}
        \dot{\mathbf x} = \mathbf v(\mathbf x,t),\quad \mathbf x(t_0) = \mathbf x_0,
    \end{equation}
    where $\mathbf v(\mathbf x,t)$ is the fluid's velocity.  
\end{definition} 

It is taken for granted that $\mathbf v$, assumed two dimensional, depends on $\mathbf x$ and $t$ sufficiently smoothly for $F_{t_0}^t$ to exist and be unique, and its derivatives of any order as needed are continuous. From this point forward, $\mathbf x = (x,y)$ is considered to vary within an open domain of $\mathbb R^2$ and $t$ within an open interval of finite length.

\begin{definition}[LAVD] 
     Let
    \begin{equation}
        \omega(\mathbf x,t) := \nabla^\perp\cdot \mathbf v(\mathbf x,t) := \partial_yv_x(\mathbf x,t) - \partial_xv_y(\mathbf x,t)
    \end{equation}
    be the fluid's vorticity.  Let  $U(t)$ be a material region, i.e, transported under flow map $F_{t_0}^t$ generated by the fluid's velocity. The \textbf{Lagrangian Averaged Vorticity Deviation} (\textbf{LAVD}) is defined by \textup{\cite{Haller-etal-16}}
    \begin{equation}
        \mathrm{LAVD}_{t_0}^t(\mathbf x_0) := \int_{t_0}^t |\omega(F_{t_0}^{t'}(\mathbf x_0),t'),t') -\bar{\omega}(t')|\,dt', 
        \label{eq:lavd}
    \end{equation}
    where 
    \begin{equation}
        \bar{\omega}(t) := \frac{\int_{U(t)} \omega(\mathbf x,t)\,d^2x}{\int_{U(t)}d^2x},
    \label{eq:bar}
    \end{equation}
    which is an average of the vorticity over $U(t)$.
\end{definition}

\begin{definition}[RCV]
    A \textbf{Rotationally Coherent Vortex} (\textbf{RCV}) over  $t \in [t_0, t_0 + T]$ is an evolving material region $V(t)\subset U(t)$, $t \in [t_0, t_0 + T]$, such that its time-$t_0$ position is enclosed by the outermost, sufficiently convex isoline of $\mathrm{LAVD}_{t_0}^{t_0+T}(\mathbf x_0)$ around a local (nondegenerate) maximum (resp., minimum) when $T > 0$ (resp., $T < 0$).
\end{definition}

More precisely, while a region $V(t)$ may encompass multiple local extrema \citep{Beron-etal-19-PNAS}, we deliberately omit such cases from our consideration to facilitate a clear definition of the center of the vortex \citep{Haller-etal-16}. Consequently, the boundary elements of these material regions $V(t)$ undergo an equivalent total material rotation relative to the mean material rotation of the entire fluid mass within the domain $U(t)$ that encompasses them. This characteristic of the boundaries predominantly constrains their filamentation to tangential directions under advection from $t_0$ to $t_0 + T$ \citep{Haller-etal-16}.

\begin{definition}[LCV]
    The boundary of a \textbf{Lagrangian Coherent Vortex} (\textbf{LCV}) \textup{\cite{Haller-Beron-13}} is the outermost member of a family of \textbf{p-loops}.  That is, limit cycles of the \textbf{p-line fields} 
    \begin{equation}
        \mathbf l_p^\pm(\mathbf x_0) :=
        \sqrt{
        \frac
        {\lambda_2(\mathbf x_0) - p^2}
        {\lambda_2(\mathbf x_0) - \lambda_1(\mathbf x_0)}
        }
        \,\boldsymbol\xi_1(x_0) 
        \pm
        \sqrt{
        \frac
        {p^2 - \lambda_1(\mathbf x_0)}
        {\lambda_2(\mathbf x_0) - \lambda_1(\mathbf x_0)}
        }
        \,\boldsymbol\xi_2(x_0),  
        \label{eq:eta}
    \end{equation}
    where $\lambda_1(\mathbf x_0) < p^2 < \lambda_2(\mathbf x_0)$.  Here, $\{\lambda_i(\mathbf x_0)\}$ and $\{\boldsymbol\xi_i(\mathbf x_0)\}$ satisfying
    \begin{equation}
        0 < \lambda_1(\mathbf x_0) =
        \frac{1}{\lambda_2(\mathbf x_0)} < 1,\quad 
        \boldsymbol\xi_i(\mathbf x_0)\cdot \boldsymbol\xi_j(\mathbf x_0) = \delta_{ij}\quad
        i,j = 1,2,
        \label{eq:eig}
    \end{equation}
    are eigenvalues and (orientationless) normalized eigenvectors, respectively, of the representation in $\mathbb R^2$ of the \textbf{right Cauchy--Green strain tensor},
    \begin{equation}
        \mathsf C_{t_0}^{t_0+T}(\mathbf x_0) := \big(\partial_{\mathbf x_0}F_{t_0}^{t_0+T}(\mathbf x_0)\big)^\top \partial_{\mathbf x_0}F_{t_0}^{t_0+T}(\mathbf x_0).  
        \label{eq:C}
    \end{equation}
\end{definition}

The $p$-line fields stationarize integrated relative stretching.  The $p$-loops undergo uniform material stretching from $t_0$ to $t_0+T$. A $1$-loop does not experience any stretching. In the incompressible case, i.e., $\nabla\cdot\mathbf v = 0$, an LCV bounded by a 1-loop is distinguished from other LCVs by the exceptional coherence of its boundary: it reassumes at $t_0+T$ the perimeter it had at $t_0$ while preserving the enclosed area. The $p$-loops can also be interpreted as \emph{null-geodesics} of the indefinite tensor field, generalizing the Green--Lagrange tensor of continuum mechanics, with representation given by $\mathsf C_{t_0}^{t_0+T}(\mathbf x_0) - p\mathsf I$ \cite{Haller-Beron-14}. Finally, RCV boundaries typically run close to LCV boundaries \cite{Andrade-etal-20}, both representing observer-independent objects. However, geodesic eddy detection demonstrates greater numerical stability compared to LAVD-based eddy detection \cite{Andrade-etal-20}. This observation, coupled with the parameter-free nature of the method, renders geodesic detection the preferred technique for material coherence analysis.

In conclusion, the term ``fluid'' used in the preceding definitions should be interpreted specifically as ocean fluid (water) to allow subsequent application of these definitions.

\subsection{Assumptions}

\begin{assumption}
    Suppose that the eBOMB system Eq.\@~(2) can be approximated by
    \begin{equation}
        \dot{\mathbf x}_i = \mathbf v_i = (u_i,v_i) := gf_0^{-1}\nabla^\perp\left.\eta\right\vert_i + \tau \big(g(1 - R)\nabla\left.\eta\right\vert_i + \mathbf F_i\big),\quad i = 1, 2, \ldots, N,
        \label{eq:eBOMB-qg}
    \end{equation}
    where $g$ is the acceleration of gravity, $f_0$ is a reference Coriolis parameter, and $\eta(\mathbf x,t)$ is the sea surface height, assumed sufficiently smooth on each of its arguments. 
\end{assumption}

This assumption is valid under the following conditions.

\begin{enumerate}
    \item The near-surface ocean flow must be in quasigeostrophic balance \cite{Pedlosky-87}, which is anticipated for mesoscale ocean flow \citep{Fu-etal-10}.  Let $\mathrm{Ro} = \mathcal U/\mathcal L|f_0| > 0$, where $\mathcal U$ and $\mathcal L$ are characteristic velocity and length scales, be a small Rossby number.  The quasigeostrophic scaling \citep{Pedlosky-87} implies that: i) the ocean velocity $\mathbf v = gf_0^{-1}\nabla^\perp\eta + O(\mathrm{Ro}^2)$, where $\eta = O(\mathrm{Ro})$, ii) $\partial_t = O(\mathrm{Ro})$, and iii) $y/a_\odot = O(\mathrm{Ro})$, implying that the Coriolis parameter $f = f_0 + O(\mathrm{Ro})$, and the geometric terms $\gamma_\odot = 1 + O(\mathrm{Ro})$ and $\tau_\odot = O(\mathrm{Ro})$.

   \item The parameter $\alpha = O(\mathrm{Ro})$, at a minimum, aligns with it being very small (a few percent) over a substantial range of $1 \le \delta < O(1) \sim 1$ permissible values, as illustrated in Fig.\@~2 of \cite{Beron-etal-19-PoF}. Specifically, with $\delta = 1.14$ as determined through optimization relative to observations in this study, we obtain $\alpha = 3.37 \times 10^{-3}$ (see Table 2). This value is very small and aligns more closely with $\alpha = O(\mathrm{Ro}^3)$ for a Rossby number typically characterizing mesoscale flow ($\mathrm{Ro} = 0.1$). It is important to note that this results in $\alpha \mathbf v_\text{W} = O(\mathrm{Ro}^3)$ for an $O(1)$ near-surface wind velocity. However, this would be inconsistent with the quasigeostrophic ocean flow assumption. Therefore, an additional condition is required.

   \item The wind velocity $\mathbf v_\text{W} = O(\mathrm{Ro}^2)$, at a minimum. This implies that the wind field over the period of interest is sufficiently weak (calm).
\end{enumerate}

It is important to highlight that, unlike Eq.\@~(2), Eq.\@~\eqref{eq:eBOMB-qg} excludes any geometric terms associated with the Earth's sphericity. Additionally, Eq.\@~(4.3) in \cite{Beron-Miron-20} unnecessarily, though insignificantly, includes the factor $1-\alpha-R$, which can be sufficiently approximated by $1-R$. (In fact, $R = 1$ when $\delta = 1$ and remains $O(1)$ within the physically relevant range $1 \le \delta < O(1) \approx 1$. In particular, for $\delta = 1.14$ one computes $R = 0.823$.) Equation (4.3) in \cite{Beron-Miron-20}, in contrast to Eq.\@~\eqref{eq:eBOMB-qg}, was derived under the assumption of Cartesian geometry from the beginning.

Assuming $\tau = O(\varepsilon)$ as mentioned in \cite{Beron-Miron-20}, Eq.\@~\eqref{eq:eBOMB-qg} is accurate with an error of $O(\varepsilon^3)$. Contrary to \cite{Beron-Miron-20}'s assertion, the corresponding Eq.\@~(4.3) also has the same error magnitude. This assumption is unnecessary, and we do not use it here, as the magnitude of $\tau$ is already determined by the Stokes number, $\mathrm{St}$, which is considered to be independently small. Hence, with $\mathrm{St}$ fixed, Eq.\@~\eqref{eq:eBOMB-qg} is valid with an error of $O(\mathrm{Ro}^2)$.

\begin{assumption}
    The elastic force $\mathbf F_i = O(\mathrm{Ro})$.
\end{assumption}

In a formal sense, ensuring that $\mathbf{F}_i \ll O(\mathrm{St}^{-1})$ is sufficient to maintain the properties of the slow manifold. By requiring the elastic force $\mathbf{F}_i$ to be of the same order as the sea surface height $\eta$, as specified by the above condition, we can achieve an attraction result.

\begin{assumption}\label{ass:kij}
    Assume that the stiffness $k_{ij}$ in Eq.\@~(3) depends smoothly on $\mathbf x_{ij} = \mathbf x_i - \mathbf x_j$, $i\neq j = 1, 2, \ldots, N$.
\end{assumption}

This assumption holds for the specific eBOMB model's $k_{ij}(|\mathbf{x}_{ij}|)$, given in Eq.\@~(4). Along with the assumed smooth dependence of $\eta(\mathbf{x}, t)$ on each of its arguments, it guarantees that the eBOMB model approximation Eq.\@~\eqref{eq:eBOMB-qg} generates a well-defined flow map.

Henceforth, we shall designate as eBOMB particles those particles that evolve in accordance with \eqref{eq:eBOMB-qg}, an approximation of Eq.\@~(2) under the specified conditions delineated previously. 

\subsection{Attraction result}

We are now ready to rigorously formulate Thm.\@~1.

\begin{theorem}[Formal statement]
    Suppose that the quasigeostrophic velocity field $\mathbf v = gf_0^{-1}\nabla^\perp\eta$, which produces the flow map $F_{t_0}^t$, contains an RCV that covers a region $V(t)$.  Let
    \begin{equation}
        \mathbf x_0^* = \argmax_{\mathbf x_0\in V(t_0)} \mathrm{LAVD}_{t_0}^{t_0+T}(\mathbf x_0),
    \end{equation}
    which naturally defines the center of the RCV. The trajectory $F_{t_0}^{t}(\mathbf x_0^*)$ of the center of the RCV locally forward attracts eBOMB particles overall over $t\in [t_0,t_0+T]$:
    \begin{enumerate}
        \item for all $k_{ij}$ when $\operatorname{sign}_{t\in [t_0,t_0+T]}\nabla^2\eta(F_{t_0}^t(\mathbf x_0^*),t) < 0$; and
        \item provided that
        \begin{equation}
            |T|\sum_{i=1}^N\sum_{j\in \operatorname{neighbor}(i)} k_{ij} > gN(1-R) \left\vert\int_{t_0}^{t_0+T} |\nabla^2\eta(F_{t_0}^t(\mathbf x_0^*),t)|\,dt\right\vert 
            \label{eq:k}
        \end{equation}
        when $\operatorname{sign}_{t\in [t_0,t_0+T]}\nabla^2\eta(F_{t_0}^t(\mathbf x_0^*),t) > 0$.
    \end{enumerate}
\end{theorem}

The consequence for \emph{Sargassum} rafts is as follows. Given that the ocean vorticity is $\omega = gf_0^{-1}\nabla^2\eta$ with an $O(\varepsilon^2)$ error, the centers of cyclonic rotationally coherent (nearly Lagrangian coherent) quasigeostrophic eddies act as finite-time forward local overall attractors for \emph{Sargassum} rafts in calm wind conditions if they are strongly connected, whereas the centers of anticyclonic eddies do so even if they are only loosely connected.

\begin{proof}
    Let $\tilde{\mathbf x} := (x_1, x_2, \dotsc, x_N, y_1, y_2, \dotsc, x_N)$ and $\tilde{\mathbf v} := (u_1, u_2,\dotsc, u_N, v_1, v_2, \dotsc, v_N)$. Then write \eqref{eq:eBOMB-qg} as
    \begin{equation}
        \frac{d\tilde{\mathbf x}}{dt} = \tilde{\mathbf v}(\tilde{\mathbf x},t).
        \label{eq:dxdt}
    \end{equation}
    Denote by $\tilde F_{t_0}^t$ the flow map corresponding to Eq.\@~\eqref{eq:dxdt} and invoke Liouville's theorem to note that a trajectory of Eq.\@~\eqref{eq:dxdt}, $\smash{\tilde F_{t_0}^t}(\tilde{\mathbf x}_0)$, is forward attracting overall over $t\in [t_0,t_0+T]$, $T>0$ (resp., $T<0$), if $\det\smash{\partial_{\tilde{\mathbf x}_0}\tilde F_{t_0}^{t_0+T}}(\tilde{\mathbf x}_0)$ is smaller (resp., larger) than unity.  Let $\varepsilon > 0$ be a small parameter.  Suppose that $|\mathbf x_i(t_0) - \mathbf x_0| = O(\varepsilon)$, $i = 1, 2, \dotsc, N$. Then, by the smooth dependence of the solutions of \eqref{eq:eBOMB-qg} on parameters, for $t\in [t_0, t_0+T]$ finite, it follows that
    \begin{equation}
        \mathbf x_i(t;\mathbf x_i(t_0),t_0) = F_{t_0}^t(\mathbf x_0^*) + O(\varepsilon),\quad i = 1, 2, \dotsc,N.
        \label{eq:xi}
    \end{equation}
    Furthermore, with an $O(\varepsilon^2)$ error, 
    \begin{subequations}
    \begin{equation}
        \det \partial_{\tilde{\mathbf x}_0}\tilde F_{t_0}^{t_0+T}(\tilde{\mathbf x}_0) = \exp\tau (\mathcal A + \mathcal B),
    \end{equation}
    where 
    \begin{align}
        \mathcal A &:= gN(1-R)\int_{t_0}^{t_0+T} \nabla^2\eta(F_{t_0}^t(\mathbf x_0^*),t)\,dt\\
        \mathcal B &:= -T\sum_{i=1}^N\sum_{j\in \operatorname{neighbor}(i)}k_{ij}.
    \end{align}
    \label{eq:AB}%
    \end{subequations}
    The proof follows by noting that $R \le 1$ and by assuming that the fluid region $U(t)$ that contains the fluid region $V(t)$ occupied by the RCV is sufficiently large, e.g.,  $U(t) = O(\varepsilon^{-1})$. Indeed, under this condition,
    \begin{equation}
        \mathcal A = gN(1-R)\operatorname{sign}\big(T\nabla^2\eta(F_{t_0}^t(\mathbf x_0^*),t)\big) \left|\int_{t_0}^{t_0+T} \left|\nabla^2\eta(F_{t_0}^t(\mathbf x_0^*),t) - \overline{\nabla^2\eta}(t)\right| \,dt\right|
    \end{equation}
    with an $O(\varepsilon^2)$ error.  The integral in the above expression is proportional to $\operatorname{LAVD}_{t_0}^{t_0+T}(\mathbf x_0^*)$, which is locally maximal in $V(t) \subset U(t)$.

    The step that remains is to show that Eqs.\@~\eqref{eq:AB} holds, for which we adapt the proof given by \citep{Beron-Miron-20} for the constant stiffness $k_{ij}$ case.  As in \citep{Beron-Miron-20}, we let $\nabla_i = \partial_{\mathbf x_i}$. Then for $\mathbf x_{ij}$-dependent $k_{ij}$, Eqs.\ (B3) and (B7) in \citep{Beron-Miron-20} read, respectively, as
    \begin{equation}
        \sum_{i=1}^N\nabla_iA_i = - \sum_{i=1}^N\sum_{j\in\operatorname{neighbor}(i)} 2k_{ij} + \mathbf x_{ij}\cdot\nabla_ik_{ij}
    \end{equation}
    and
    \begin{equation}
        \sum_{i=1}^N\nabla_iB_i = \sum_{i=1}^N\sum_{j\in\operatorname{neighbor}(i)} \frac{L}{|\mathbf x_{ij}|} \left(k_{ij} + \mathbf x_{ij}\cdot\nabla_ik_{ij}\right),
    \end{equation}
   upon identifying $L$ with $\ell_{ij}$. Equation (B9) in \citep{Beron-Miron-20} remains the same since the terms involving $\nabla_ik_{ij}$ cancel out.  That is, with an $O(\varepsilon^2)$ error,
   \begin{equation}
       \operatorname{trace}\partial_{\tilde{\mathbf x}}\tilde{\mathbf v}(\tilde F_{t_0}^{t}(\tilde{\mathbf x}_0),t) = \tau gN(1-R)\nabla^2\eta(F_{t_0}^t(\mathbf x_0^*),t) - \tau\sum_{i=1}^N\sum_{j\in \operatorname{neigh}(i)} k_{ij},
   \end{equation}
   from which Eq.\@~\eqref{eq:AB} follow since $\det\partial_{\tilde{\mathbf x}_0}\tilde F_{t_0}^{t}(\tilde{\mathbf x}_0) = \int_{t_0}^{t}\operatorname{trace} \partial_{\tilde{\mathbf x}} \tilde{\mathbf v}(\tilde F_{t_0}^{t'}(\tilde{\mathbf x}_0),t')\,dt'$.
\end{proof}

\newpage

An important final observation is that maximization of $\mathrm{LAVD}_{t_0}^{t_0+T}(\mathbf x_0)$ at $\mathbf x_0^*$ guarantees that the trajectory of the fluid particle starting at $\mathbf x_0^*$ forward attracts eBOMB particles overall over $t\in [t_0,t_0+T]$ at the \emph{strongest} rate among fluid particles trajectories $O(\varepsilon)\, C^1$-close to $F_{t_0}^t(\mathbf x_0^*)$. In fact, this implies that $\det\smash{\partial_{\tilde{\mathbf x}_0}\tilde F_{t_0}^{t_0+T}}(\tilde{\mathbf x}_0) - 1 > 0$ (resp., $\det\smash{\partial_{\tilde{\mathbf x}_0}\tilde F_{t_0}^{t_0+T}}(\tilde{\mathbf x}_0) - 1 < 0$) is the \emph{largest} (resp., the \emph{smallest}) when $T<0$ (resp., $T>0$) among $\tilde{\mathbf x}_0$ such that $|\mathbf x_i(t_0) - \mathbf x_0^*| = O(\varepsilon)$, $i = 1, 2, \dotsc, N$.

\section{Additional comparisons}
\label{sec:additional-comparisons}

We present global time series comparisons between the \ebomb{} and leeway models for three additional metrics. The first metric is the $L^1$ metric, straightforwardly defined between two distributions $P_1, P_2$ supported on sample space $\mathcal{X}$ as
\begin{equation} \label{eq:L1-defn}
    \parallel P_1 - P_2 \parallel_1 = \sum_{x \in \mathcal{X}} |P_1(x) - P_2(x)|.
\end{equation}
The second metric is the Mean-Squared Error (MSE) defined as
\begin{equation} \label{eq:mse-defn}
    \text{MSE}(P_1, P_2) = \frac{1}{|\mathcal{X}|}\sum_{x \in \mathcal{X}} (P_1(x) - P_2(x))^2.
\end{equation}
The third metric is the Structural Similarity Index (SSIM) \cite{wang2004image}. This is a well-known measurement of the similarity between two images, although a full definition is beyond the scope of this paper. We note that it is strictly speaking not a distance function since it can be negative, although it still provides a useful comparison of the perceptual similarity between heatmaps.

\begin{figure}
    \centering
    \includegraphics[width = .75\textwidth]{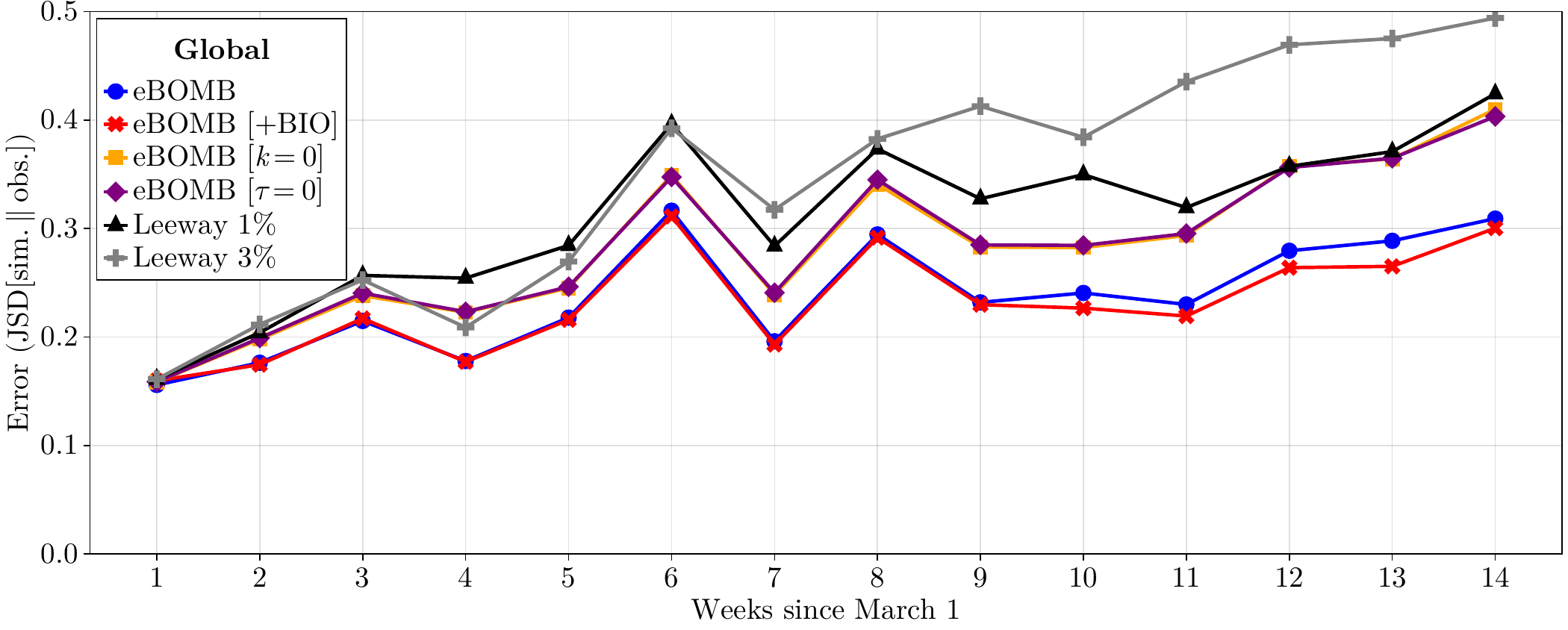}
    \caption{As Fig.\@~4 with additional comparisons between springless and intertialess \ebomb{}.}
    \label{fig:tsg-jsd}
\end{figure}

\begin{figure}
    \centering
    \includegraphics[width = .75\textwidth]{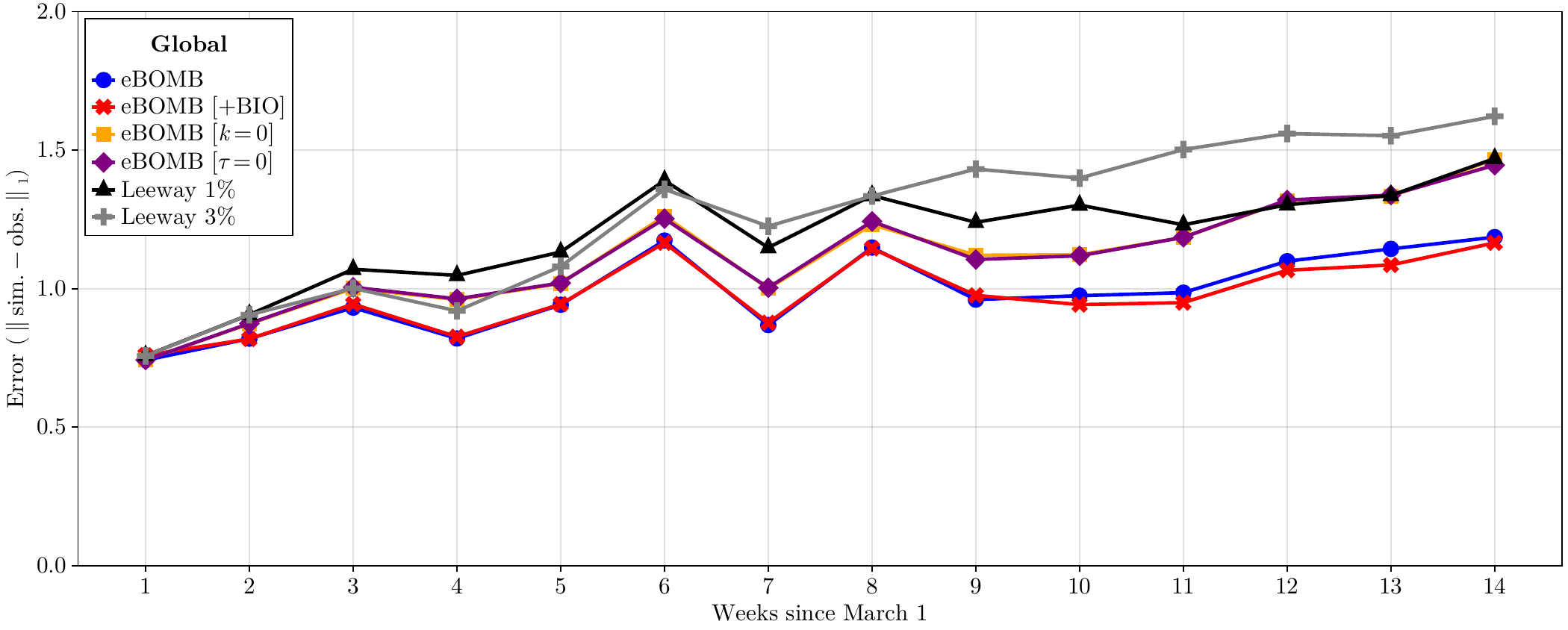}
    \caption{As Fig.\@~\ref{fig:tsg-jsd} for the $L^1$ metric.}
    \label{fig:tsg-l1}
\end{figure}

\begin{figure}
    \centering
    \includegraphics[width = .75\textwidth]{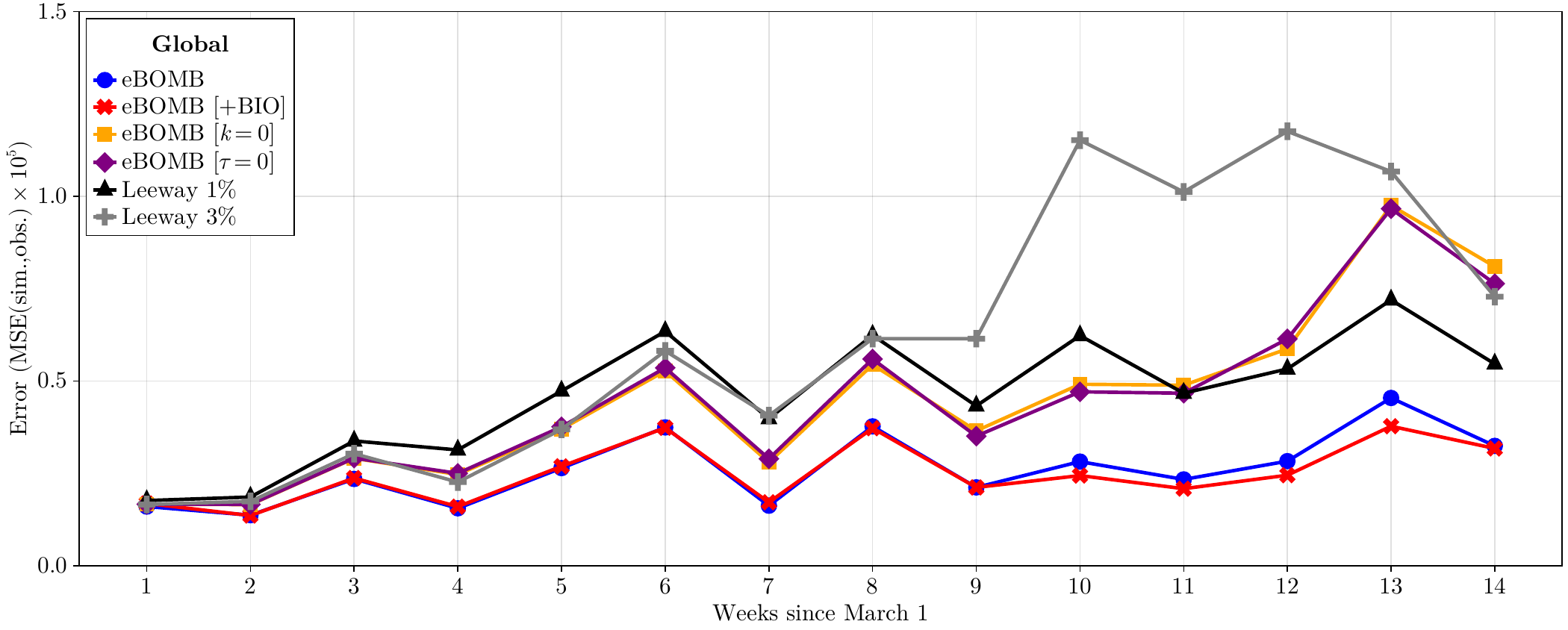}
    \caption{As Fig.\@~\ref{fig:tsg-jsd} for the MSE metric.}
    \label{fig:tsg-mse}
\end{figure}

\begin{figure}
    \centering
    \includegraphics[width = .75\textwidth]{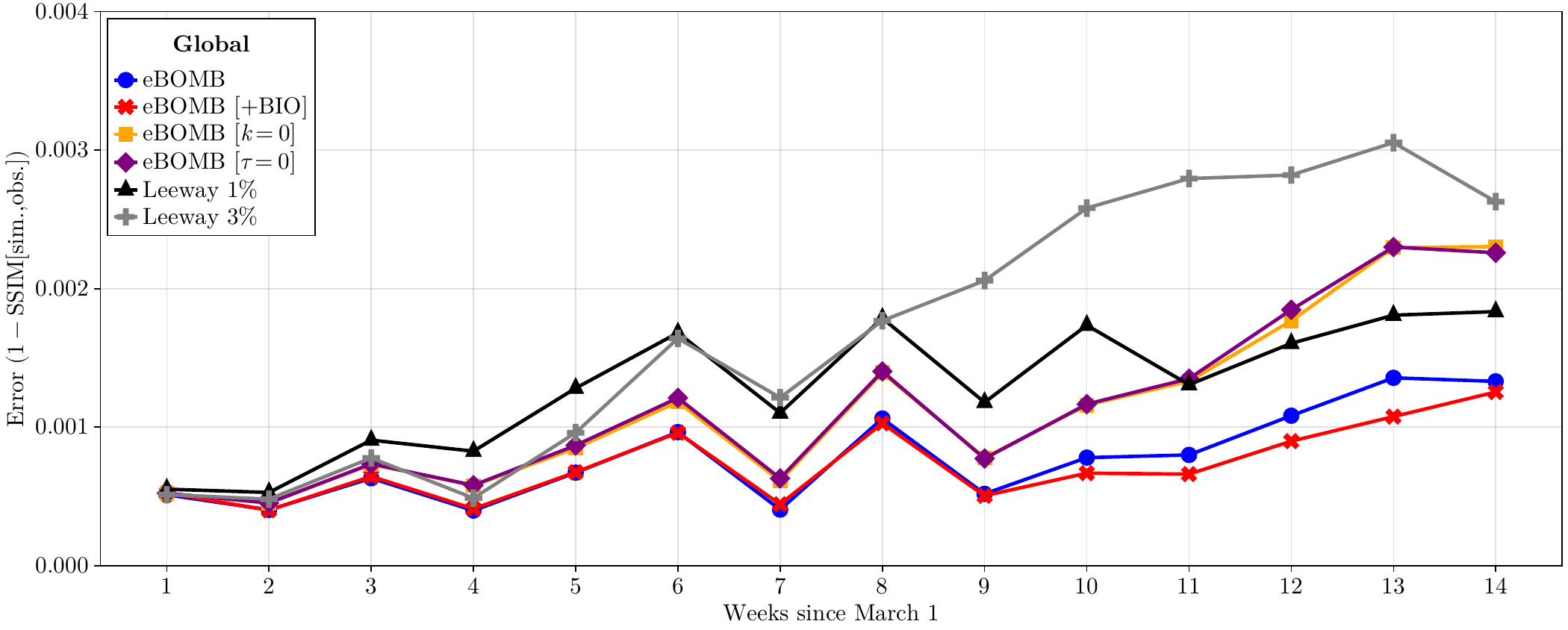}
    \caption{As Fig.\@~\ref{fig:tsg-jsd} for the SSIM metric.}
    \label{fig:tsg-ssim}
\end{figure}

\bibliographystyle{alpha}
%\bibliography{bibs/fot, bibs/fot2, bibs/data-sources}
\newcommand{\etalchar}[1]{$^{#1}$}

\end{document}